\documentclass{natureprintstyle}
\usepackage{graphicx}
\usepackage{dcolumn}
\usepackage{bm}
\usepackage{amsmath}
\usepackage{amssymb}
\usepackage{float}
\usepackage{lipsum}
\usepackage{color}
\usepackage{times}
\usepackage{textcomp}
\usepackage{latexsym}
\usepackage[hidelinks]{hyperref}
\usepackage{mathtools}
\usepackage{url}

\usepackage[utf8]{inputenc}

\newcommand{\tn}{\textnormal}

\begin{document}

\title{Detecting new fundamental fields with LISA}

\author{Andrea Maselli,$^{1,2}$ 
Nicola Franchini,$^{3,4}$
Leonardo Gualtieri,$^{5,6}$ 
Thomas P. Sotiriou,$^{7,8}$ 
Susanna Barsanti,$^{5,6}$
Paolo Pani$^{5,6}$} % authors

\maketitle

\begin{affiliations}
\item %AM
Gran Sasso Science Institute (GSSI), I-67100 L’Aquila, Italy.
\item %AM
INFN, Laboratori Nazionali del Gran Sasso, I-67100 Assergi, Italy.
\item % NF
SISSA, Via Bonomea 265, 34136 Trieste, Italy and INFN Sezione di Trieste.
\item %NF 
IFPU - Institute for Fundamental Physics of the Universe, Via Beirut 2, 34014 Trieste, Italy.
\item % SB, LG, PN
Dipartimento di Fisica, ``Sapienza'' Universit\'a di Roma, Piazzale Aldo Moro 5, 00185, Roma, Italy.
\item % SB, LG, PN
Sezione INFN Roma1, Piazzale Aldo Moro 5, 00185, Roma, Italy.
\item Nottingham Centre of Gravity, University of Nottingham, University Park, Nottingham, NG7 2RD, UK.
\item % T.S.
School of Mathematical Sciences \& School of Physics and Astronomy, University of Nottingham, University Park, Nottingham, NG7 2RD, UK.
\end{affiliations}

\smallskip

\maketitle

\begin{abstract}
The Laser Interferometer Space Antenna,\cite{2017arXiv170200786A} LISA, 
will detect gravitational wave signals from Extreme Mass Ratio 
Inspirals,\cite{Babak:2017tow} where a stellar mass compact object  orbits a supermassive black hole and eventually plunges into it. Here we report on LISA's capability to detect whether the smaller compact object in an Extreme Mass Ratio Inspiral is endowed with a  scalar field,\cite{Berti:2015itd,Barausse:2020rsu} and to measure its scalar charge --- a dimensionless quantity that acts as a measure of how much scalar field the object carries. By direct comparison of signals, we show that LISA will be able to detect and measure the scalar charge with an accuracy of the order of percent, which is an unprecedented level of precision. This result is independent of the origin of the scalar field and of the structure and other properties of the small compact object, so it can be seen as a generic assessment of LISA's capabilities to detect new fundamental fields. 
\end{abstract}

Such fields, and scalars in particular, are ubiquitous in extensions of General Relativity (GR) or the Standard Model. Depending on the model, they might be low-energy classical remnants of Quantum Gravity, they might address the hierarchy problem or other naturalness problems in the Standard Model, they may explain the accelerated expansion of the universe, or account for dark matter.\cite{Berti:2015itd,Copeland:2006wr,Essig:2013lka,Hui:2016ltb,Barack:2018yly,Barausse:2020rsu} So far, searches for new fundamental fields in the vicinity of astrophysical objects with weak gravitational fields, or small spacetime curvatures, have yielded strong constraints rather than detections.  However,  gravitational wave (GW) observations are starting to probe for the first time strong gravitational fields, or larger spacetime curvature, and there is hope that new fields will be detected in this regime.

This hope is not based purely on pragmatism related to the strong constraint from weak gravity experiments. From a theoretical standpoint, it is quite reasonable and intuitive to expect that deviations from GR, or in the interactions between gravity and new fields that are part of some extension of the Standard Model, will be more prominent at larger curvatures. Indeed, this is the behaviour one tends to find in models where black holes (BHs) or compact stars are known to exhibit different structure than their GR counterparts.\cite{Damour:1993hw,Kanti:1995vq,Yunes:2011we,Sotiriou:2013qea} Moreover, there are specific effects, such as superradiance\,\cite{zeldovich1,Brito:2015oca,Arvanitaki:2009fg,Cardoso:2011xi} or spontaneous scalarization,\cite{Damour:1993hw,Silva:2017uqg,Doneva:2017bvd} that can render new fields detectable only in the vicinity of compact objects. 

In this context, Extreme Mass Ratio Inspirals (EMRIs) are perhaps somewhat special. Although both of their constituents are compact objects and one tends to associate a strong gravitational field to both of them, they have very different masses and, hence, very different characteristic curvatures. The central (primary) object is expected to be a supermassive BH (with  mass above $M\sim 10^5$ solar masses) and the curvature near the horizon will thus be very small (it scales as one over the mass squared). The small (secondary) object, irrespective of what it is, has much larger characteristic curvature. Combined with the above, this implies that, at least for some sizeable subset of EMRIs, the primary object can be taken to be a Kerr BH to a high degree of precision and any potential deviation will only affect the smaller secondary object (see the {\it Methods} below).\cite{Maselli:2020zgv}
This observation introduces an important simplification both in the modelling of the signal in presence of new fundamental fields, and in the conceptual understanding of how deviations in the signal arise. 

Let us consider an EMRI in which the secondary object (which can be a neutron star, a black hole or something more exotic) moves in the 
spacetime generated by the primary body and is endowed with a massless scalar field configuration.   In a local reference frame 
$\{{\tilde x}^\mu\}$ centered on the secondary body the scalar 
field has the form
\begin{equation}
\varphi=\varphi_0+\frac{\mu\,d}{\tilde r}+O\left(\frac{\mu^2}{{\tilde r}^2}\right)\label{sol:scalhom}
\end{equation}
where $\tilde{r}$ is the radial coordinate, 
$\mu<< M$ is the mass of the secondary body, $d$ is its 
dimensionless scalar charge, $\varphi_0$ is the asymptotic 
value of the scalar field, and we are using geometric 
units $G = c = 1$, with $G$ and $c$ being the gravitational 
constant and the speed of light, respectively.
As the secondary 
body orbits the supermassive Kerr BH, it will emit GWs.
Its motion can be 
approximated by that of a point particle moving in the ``exterior'' Kerr 
spacetime, described by a set of global coordinates $\{x^\mu\}$ (e.g. the Boyer-Lindquist coordinates, see Methods).
In this frame the worldline is described the functions $y^\mu_{\rm p}(\lambda)$, where $\lambda$ is the affine parameter.
The flux for the  polarizations of the emitted GWs
will be virtually unaffected by the fact that it carries a scalar 
charge,\cite{Maselli:2020zgv} as will be discussed in more detail in the {\em Methods}. It can hence be computed using  state-of-the-art EMRI models. However, the secondary body is  accelerating due to gravity and accelerating charges emit radiation. Hence, on top of the standard GW emission, there will be also scalar radiation throughout the inspiral. 

Due to its scalar charge, the secondary body acts as a source of the scalar field equation. In the setup described above, the latter takes the form
\begin{equation}\label{eq:sourcescal}
     \square\varphi=-4\pi d\,\mu\int \frac{\delta^{(4)}(x-y_{p}(\lambda))}{\sqrt{-g}}d\lambda\ ,
\end{equation}
where $\square$ is the D'Alambertian operator for the Kerr 
metric, and $g$ is the metric determinant (see Methods).
This equation can be used to compute the the scalar field flux, $\dot E_{\rm scal}$, which is proportional to the scalar charge squared.
The orbital energy of the particle decreases due to the total energy emission:
\begin{equation}
    {\dot E}_{\rm orb}=-{\dot E}_{\rm grav}-{\dot E}_{\rm scal}\ ,
\end{equation}
where $\dot{E}_\textnormal{grav}$ is the gravitational flux. 
The scalar flux affects the orbital evolution of the secondary
body. Since the GW phase is determined by the orbital evolution, the scalar field emission contributes to the GW phase of the EMRI. LISA is expected to constrain the source parameters from the EMRI inspiral evolution with exquisite precision (see also the Supplementary material for further details).
Thus, it will also be able to measure the scalar charge of the 
secondary body, if present.

Hereafter, we study prototype EMRIs in which the primary body is a supermassive BH of $M=10^6M_\odot$ and dimensionless spin $\chi=J/M^2=0.9$, where $J$ is the BH's angular momentum. We take  the secondary body to be a compact object with mass $\mu=10M_\odot$. We track the evolution of a binary on a circular equatorial orbit for one year, before the plunge, i.e., we choose the initial radial position $r_0$ such that, after a year the secondary body is within a distance of $0.1 M$ from the innermost stable circular orbit (ISCO). Further details on the source parameters are discussed in the {\it Methods} Section. 
We shall consider values of the signal-to-noise ratio (SNR) for LISA detection of EMRIs ranging from $30$ to few hundreds, reflecting conservative or more optimistic expectations\cite{Babak:2017tow} based on rather uncertain event-rate estimates for EMRIs.

A preliminary assessment  of the detectability of the scalar charge can be made using the evolution of the phase of the GW signal. 
The fact that the phase difference between binaries with and without scalar charge exceeds a certain threshold is an indication that the scalar charge should be 
detectable by LISA after twelve months of observation. 
We have calculated the phase difference and it is indeed above threshold even for values of the scalar charge as small as $d\sim5\times10^{-3}$. More details are given in {\it Methods}.

A more quantitative analysis on LISA's ability to detect a scalar charge is given in Figure~1,  which shows the {\it faithfulness} ${\cal F}$ between  two GW signals emitted by binaries with and without the charge. The faithfulness (see the {\it Methods} below for a precise definition) provides an estimate of how much two signals differ, weighted by the noise spectral density of LISA. Given the SNR $\rho$ of a signal, values of ${\cal F}$ smaller than $\sim 1-D/(2\rho^2)$, with $D$ dimension of the model ($\sim10$), indicate that the two waveforms are significantly different and don't provide a {\it faithful} description of one another.\cite{Flanagan:1997kp,Lindblom:2008cm,Chatziioannou:2017tdw} For $\rho=30$ this requirement translates into ${\cal F}\lesssim 0.988$. 
Figure~1 shows the values of ${\cal F}$ for the chose prototype of binary configuration. After one year the faithfulness is always smaller than the threshold set by $\rho=30$, even for scalar charges as small as $d\gtrsim 0.01$. For the same configuration, on a period of just six months before the plunge, the faithfulness is below the threshold already for charges $d\gtrsim 0.05$.

The analysis carried out so far highlights two important aspects: (i) the scalar charge provides a significant shift in the phase of the GW signal emitted by EMRIs, (ii) the dephasing induces a mismatch in the template with respect to the zero-charge case, which can potentially lead to a severe loss of events and to a bias in the estimation of the waveform parameters.\cite{Lindblom:2008cm} This suggests that one year of LISA observations of EMRIs may be able to reveal the presence of a scalar charge as small as $d\sim0.005-0.01$.

%%%%%%%%%%%%%%%%%%%%%%%%%%%%%%%%%%%%%%%%%%%%%%%%%%%%%%%%%%%%%%%%
\begin{figure}
\includegraphics[width=\columnwidth]{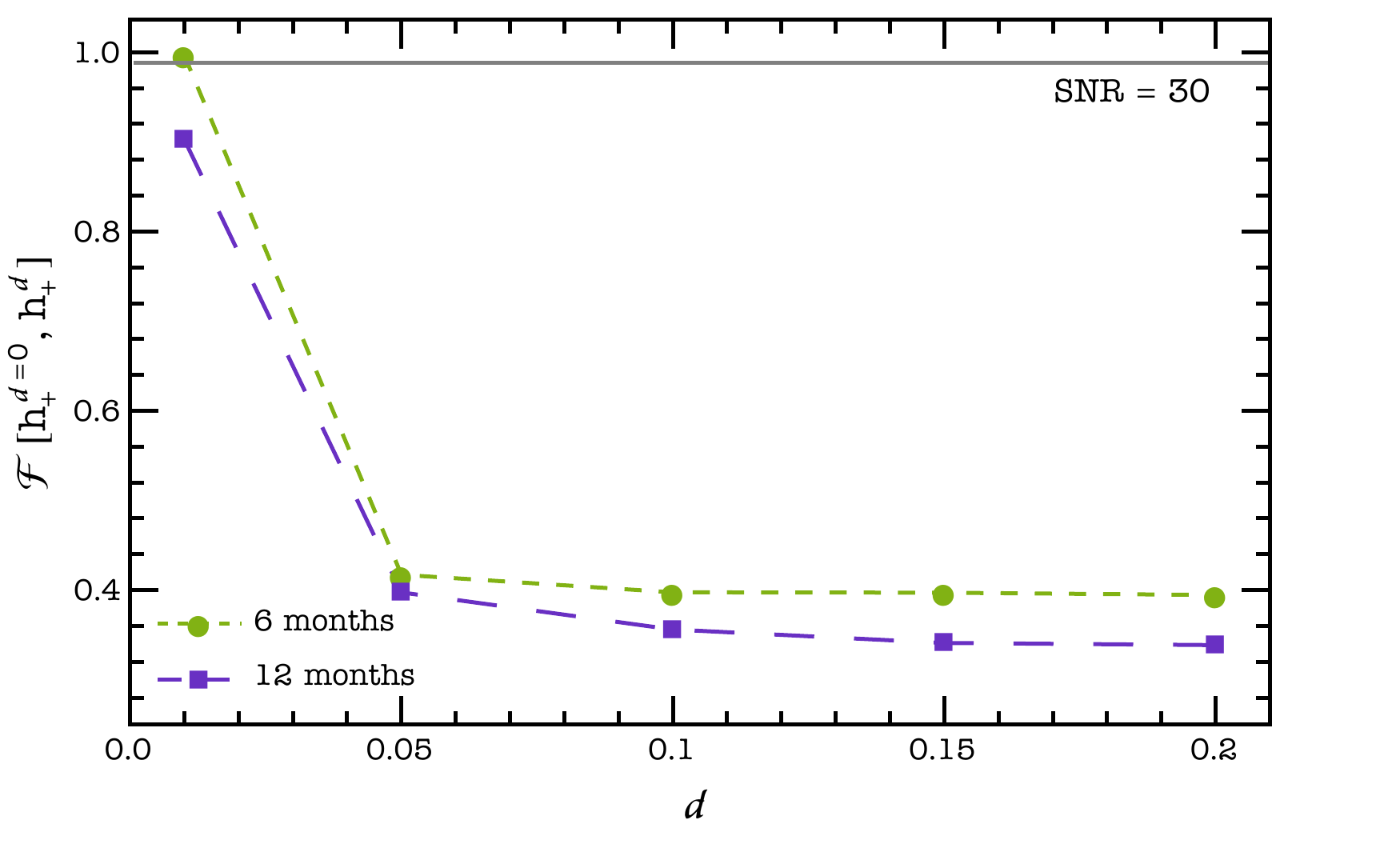}
 \caption{
 \textbf{Faithfulness between the GW plus polarization computed with and 
 	without the scalar charge, as a function of the latter and for different signal durations.} 
 The signal duration is measured in months ($6$ or $12$) before the plunge. The horizontal  line identifies the threshold of distinguishability, ${\cal F}\lesssim 0.988$, set up by SNR of $30$.}
 \label{fig:faithfulness}
\end{figure}
%%%%%%%%%%%%%%%%%%%%%%%%%%%%%%%%%%%%%%%%%%%%%%%%%%%%%%%%%%%%%%%%
\begin{figure}
 \includegraphics[width=\columnwidth]{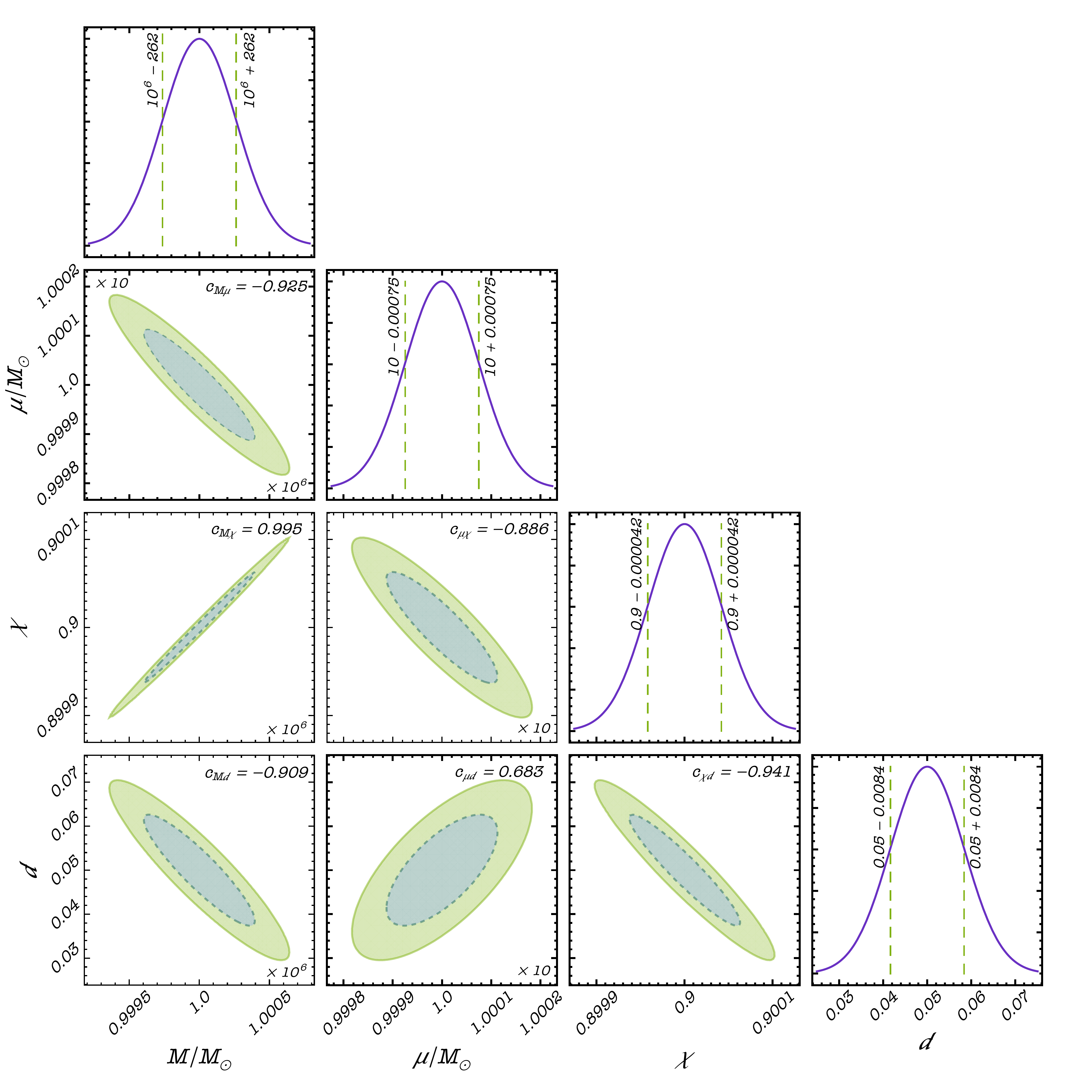}
 \caption{
  \textbf{Corner plot for the probability distribution of the masses, primary spin and secondary charge,  inferred after one year of observation with LISA.} 
 We consider a binary system with $d=0.05$ and SNR of $150$. 
 Diagonal (off-diagonal) boxes refer to marginalized (joint) distributions.
 Vertical lines show the 1-$\sigma$ interval for each waveform parameters. Colored contours within the joint distributions correspond to $68\%$ and $95\%$ probability confidence intervals. In the top right corner of each off-diagonal panel we show the  correlation coefficient $c_{ij}$ between the parameters 
 $(i,j)$ of the joint distribution.}
 \label{fig:errors_spin9}
\end{figure}
%%%%%%%%%%%%%%%%%%%%%%%%%%%%%%%%%%%%%%%%%%%%%%%%%%%%%%%%%%%%%%%%
\begin{figure}
 \includegraphics[width=\columnwidth]{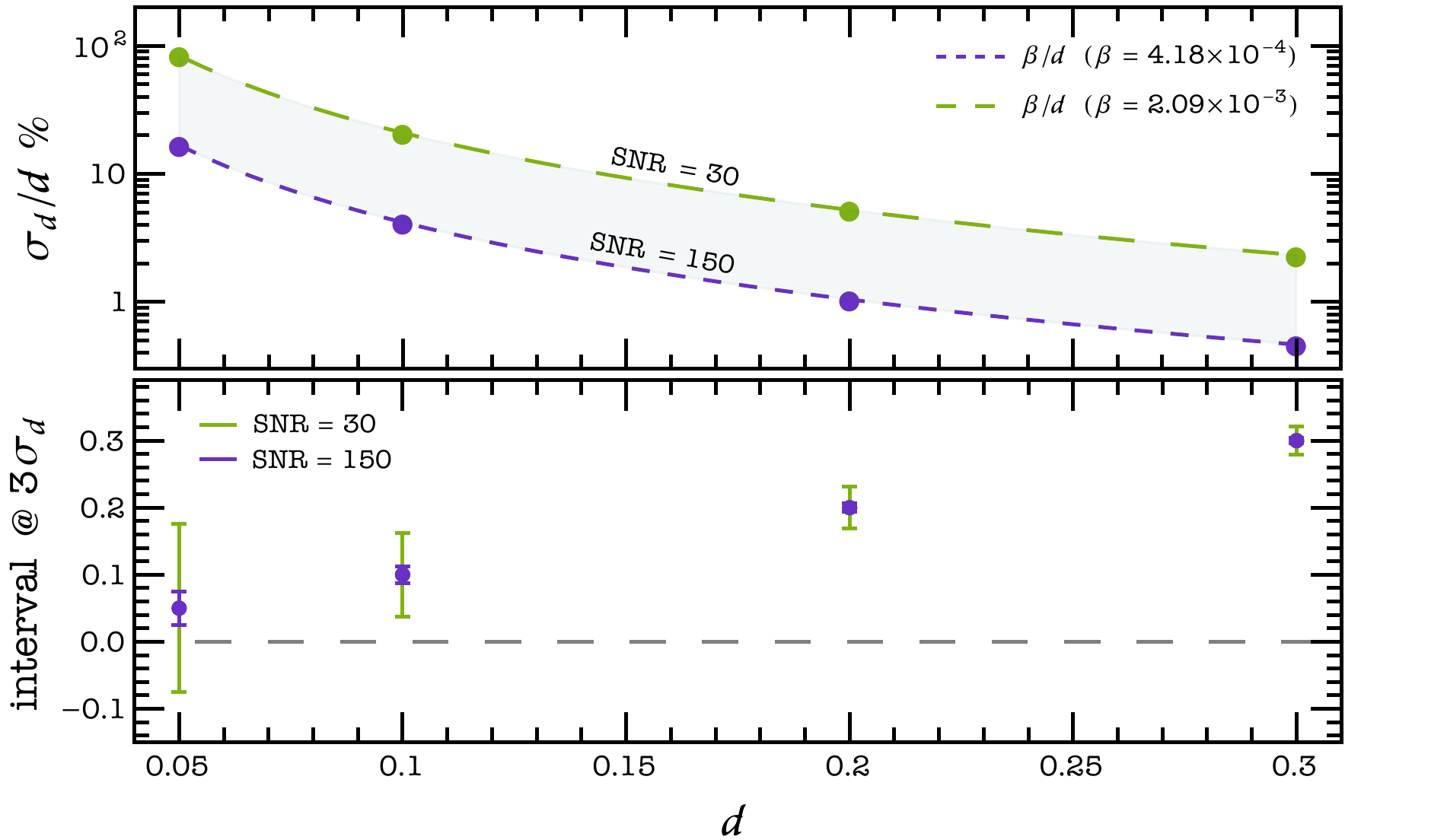}
 \caption{\textbf{Uncertainties on the scalar charge for 
 		prototype EMRIs observed with different SNR after one year of observation by LISA.}
 	(Top panel) Relative error on the scalar charge as 
 	a function of $d$ for EMRIs SNR of 30 and 150. (Bottom panel) 3-$\sigma$ interval around the true values 
 	of the scalar charge inferred from LISA observations with the two 
 	values of the SNR also considered in the top panel. 
 	Dashed curves in the top panel refer to the analytic fit $\sigma_d=\beta/d$ 
 	with $\beta\simeq4.18\times 10^{-4}$ and  $\beta\simeq2.09\times 10^{-3}$ 
 	for  SNR of 150 and 30, respectively.} \label{fig:errors}
\end{figure}
%%%%%%%%%%%%%%%%%%%%%%%%%%%%%%%%%%%%%%%%%%%%%%%%%%%%%%%%%%%%%%%%

The dephasing and the faithfulness, however, don't take fully into account possible degeneracies among the waveform parameters, which may jeopardize our ability to constrain the scalar charge. A more sophisticated study requires a joint investigation of the full parameter space, which includes correlations among $d$ and other quantities characterizing the EMRI GW emission. In the following we perform such an analysis, assessing the capability of LISA to perform an actual {\it measurement} of the scalar charge. 

Figure~2 shows the probability distribution obtained using a Fisher matrix approach (see {\it Methods}) for the component masses, the spin of the primary, and the scalar charge of the secondary, for EMRIs observed one year before the plunge with $d=0.05$ and SNR of $150$.  
This analysis shows that a single detection can provide a measurement of the scalar charge with a relative error smaller than $10\%$, with a probability distribution that does not have any support on $d=0$ at more than 3-$\sigma$. 
Off-diagonal panels, yielding $68\%$ and $98\%$ joint probability confidence intervals between the source parameters, also show that the charge is highly correlated with the secondary mass and anti-correlated with the spin parameter and the mass of the primary.

Figure~3 shows the error in the scalar charge as a function of the scalar charge itself, for EMRIs detected by LISA with SNR ranging from $30$ to $150$. The errors on  $d$ can also be accurately fitted with a simple law of the form $\sigma^\tn{fit}_d=\beta/d$, where $\beta\simeq2.09 \times 10^{-3}$ ($\beta\simeq4.18 \times 10^{-4}$) for SNR of $30$ ($150$). In the top panel we show the relative error $\sigma_d/d$ and the analytical fit; in the bottom panel we show the 3-$\sigma_d$ intervals around the injected values of the scalar charge.

Our analysis shows that one year of EMRI observation can pinpoint a scalar charge smaller than $\sim0.3$ with percent accuracy. For an SNR of $30$ a charge $d\sim0.1$ could be constrained to consistently exclude the value $d=0$. For the louder signals we consider, LISA could constrain a  scalar charge as small as $d\sim0.05$ to be inconsistent with zero at 3-$\sigma$ confidence level.

Detecting and measuring the scalar charge of a compact object would be of enormous importance,  as first evidence of new physics, regardless of the origin of the charge or the nature of the compact object. Indeed, so far our analysis and results have been theory-agnostic. However, it is worth pointing out that in many cases the scalar charge is uniquely determined by theoretical parameters that mark deviations from GR or the Standard Model. In such cases, a measurement of the scalar charge can be used to measure these parameters. LISA will provide impressive precision for that. 

Let us demonstrate this point using a simple but characteristic example. Assume that the secondary body is a black hole and the scalar field is massless (shift-symmetric). No-hair theorems then dictate that there cannot be a scalar charge unless the scalar field couples to the Gauss--Bonnet invariant, $R_{\rm GB}=R^2-4R_{\mu\nu}R^{\mu\nu}+R_{\mu\nu\alpha\beta}R^{\mu\nu\alpha\beta}$ 
(where $R_{\mu\nu\alpha\beta},R_{\mu\nu}$ 
are the Riemann and the Ricci tensor, respectively, and 
$R$ is the Ricci scalar), as follows, $\alpha\,\varphi R_{\rm GB}$, where $\alpha$ is the new coupling constant.\cite{Sotiriou:2013qea} In this case, the relation between  $\alpha$ and the scalar charge $d$ of a BH has the simple form  $\alpha\simeq 2\mathit{d} \mu ^2-73 \mathit{d}^3 \mu ^2/240$\cite{Julie:2019sab}. 

To study the constraints on $\alpha$ from LISA observations, we draw $N=10^5$ samples of $(\mu,d)_{i=1,\ldots N}$ from the joint probability distribution of the secondary mass and scalar charge obtained from the Fisher analysis. We then compute $N$ values of $\alpha$ building the corresponding probability density functions. 
Figure~4 shows ${\cal P}(\sqrt{\alpha})$ for 
our prototype EMRIs, for $d=0.05$ and $d=0.2$. Vertical lines in each panel 
identify the $90\%$ confidence intervals of the coupling constant. 
Even for $d=0.05$ the probability distribution does not have support on $\alpha=0$.
This analysis demonstrates that, in theories where the scalar charge is determined by theoretical parameters, EMRI observations by LISA can be used to measure these parameters with unprecedented accuracy.

\begin{figure}
 \includegraphics[width=\columnwidth]{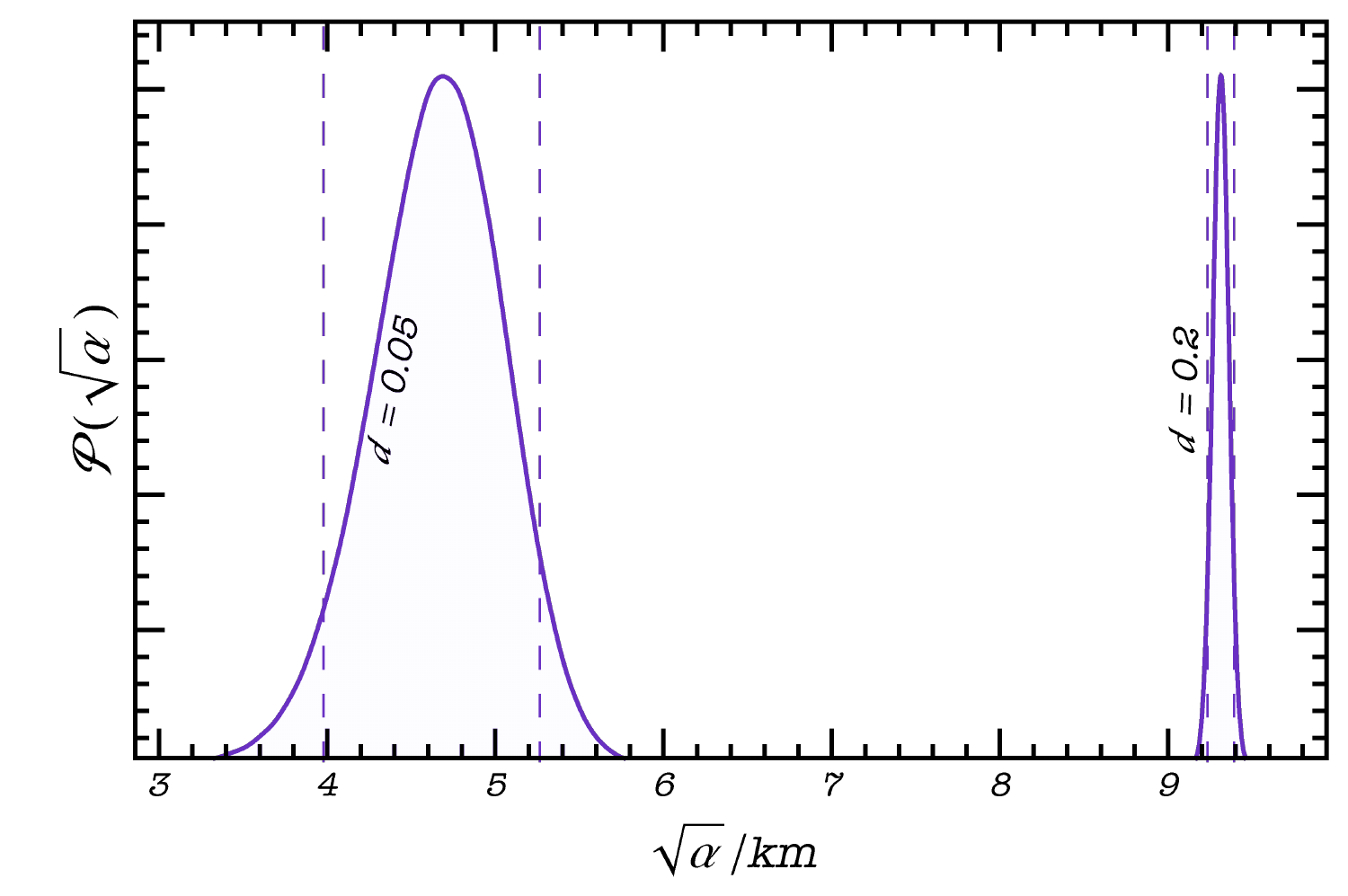}
 \caption{\textbf{Probability distribution of the dimensionful coupling constant 
 		of shift-symmetric Gauss Bonnet gravity inferred from constraints made 
 		by LISA.} Results refer to EMRIs with $d=(0.05,0.2)$ and SNR of 150. 
 	Vertical lines identify $90\%$ intervals around the injected values of 
 	the scalar charge, $\sqrt{\alpha}=4.67^{+0.59}_{-0.69}$ km and 
 	$\sqrt{\alpha}= 9.313^{+0.079}_{-0.080}$ km for $d=0.05$ and $d=0.2$, 
 	respectively (our normalization for $\alpha$ is different from the 
 	one used in some of the literature \cite{Perkins:2021mhb,Barausse:2016eii}). The height of the ${\cal P}(\sqrt{\alpha})$ distribution 
 	has been rescaled to unity.}\label{fig:errors_sGB}
\end{figure}

In summary, our results demonstrate that EMRI observations by LISA will be 
able to detect and potentially measure scalar charges to exquisite accuracy. 
Our analysis and results are independent of the origin of the charge, and are hence theory-agnostic. We have also shown that a further analysis can allow one to measure the coupling parameters for specific theories. 

This is the first attempt to perform a rigorous estimation of the measurability of beyond-GR effects with EMRIs. The EMRI template we developed here is only the starting point of a more refined analysis, which is necessary to assess LISA's full potential to detect fundamental fields and new physics beyond GR.
A number of improvements to this template would need to be made before it can be used for data analysis. These include using more generic (eccentric, non-equatorial, 
inclined) orbits,\cite{Warburton:2011hp,Warburton:2014bya,Nasipak:2019hxh} taking into account finite size effects and a full self-force treatment, or considering further beyond-GR and Standard Model effects, such as dark matter spikes or superradiant clouds. Environmental effects, as produced by accretion disks, should also be considered, in order to study possible degeneracies with the scalar emission. However, such effects would in general carry a specific frequency content, allowing to disentangle or partially alleviate correlations.\cite{Cardoso:2019rou} On the data analysis side, the next step is to perform Bayesian inference.\cite{Katz:2021yft} Although producing more realistic templates is a major technical challenge, we expect that such templates would break degeneracies between source parameters, and could potentially enhance detection capabilities. Finally, data gaps, i.e. interruptions in the interferometric measurements due to various astrophysical and instrumental factors, may provide important challenges to be overcome. Studies in this context are currently underway, showing how Bayesian methods can be able to cure this problem and improve detection efficiency.\cite{Baghi:2019eqo}

\newpage

\section*{Methods}
\noindent
%-------------------------------------------------------------
\textbf{Theoretical setup.} In this article we are assuming that the primary BH object is described, with good accuracy, by the Kerr metric, and that the particle with scalar charge $d$ sources the scalar field through Eq.~\eqref{eq:sourcescal}. This is indeed the case for a large class of theories.\cite{Maselli:2020zgv} In particular, it is quite straightforward to see that it holds in theories that are covered by no-hair theorems.\cite{Chase:1970e,Bekenstein:1995un, Hawking:1972qk, Sotiriou:2011dz,Hui:2012qt,Sotiriou:2013qea,Cardoso:2016ryw,Silva:2017uqg,Sotiriou:2014pfa} as the latter dictate that stationary BHs are described by the Kerr solution. One could also imagine a theory in which both Kerr and non-Kerr BH exist, in which case our approach would still be clearly applicable to at least some fraction of EMRIs and this could be more than enough to lead to a detection. However, although less obvious, it is important to stress that our approach is also valid for a large class of theories known to violate no-hair theorems.\cite{Maselli:2020zgv}

Consider a very general action of the form
\begin{equation}
  S[g_{\mu\nu},\varphi,\Psi]=S_0[g_{\mu\nu},\varphi]+\alpha S_c[g_{\mu\nu},\varphi]+S_{\rm m}[g_{\mu\nu},\varphi,\Psi]\,,\label{action}
\end{equation}
where
\begin{equation}
  \label{S_0}
S_0= \int d^4x \frac{\sqrt{-g}}{2\kappa}\left(R-\frac{1}{2}\partial_\mu\varphi\partial^\mu\varphi\right)\,,
\end{equation}
$\kappa=8\pi$, $R$ is the Ricci scalar, $\varphi$ is a massless scalar field, and $\alpha$ is a coupling constant with dimensions $[\alpha]=(\rm{mass})^n$ ($n>0$). Our formalism is not expected to apply to scalar fields with a non-vanishing mass $\mu_s$, which may be relevant for example as dark matter candidates.\cite{Yunes:2011aa,Hannuksela:2018izj,Annulli:2020ilw}
However, in this set up the scalar radiation is in general strongly suppressed. Indeed, for the typical velocities featured by EMRIs at the end of the inspiral the scalar flux is not quenched only for $\mu\lesssim 10^{-2}/M$ (or equivalently $\mu \lesssim 10^{-18}\,\text{eV}\left(10^6 M_\odot/M\right)$).
The term $\alpha S_c$ in Eq.~\eqref{action} describes non-minimal couplings between the metric tensor $g_{\mu\nu}$ and the scalar field, while $S_{\rm m}$ is the action of the matter fields $\Psi$.

Since in an EMRI the secondary body, with mass $\mu$, inspirals onto the primary body, a BH with mass $M\gg \mu$, we can use the so-called ``skeletonized approach'',\cite{Eardley:1975sc,Damour:1992we,Julie:2017ucp,Julie:2017rpw} in which the secondary body is treated as a point particle, characterized by a scalar function $m(\varphi)$ that depends on the value of the scalar field at the location of the particle. Strictly speaking, $m(\varphi)$ is evaluated at $\varphi=\varphi_0$ (see Eq.~\eqref{sol:scalhom}), i.e. near the particle in the lengthscale of the exterior spacetime, far away from it in the scale of the particle itself.
Finite-size effects appear at higher order compared to the leading, dissipative contribution we consider, and thus can be neglected.\cite{Steinhoff:2012rw}
This is an accurate approximation because the gravitational field of the secondary body is large only within a world-tube whose radius is much smaller than the characteristic length of the ``exterior'' spacetime, i.e., the spacetime generated by the primary body; thus, this world-tube can be treated as a world-line $y_{\rm p}^\mu(\lambda)$ in the exterior spacetime. By integrating the matter action $S_{\rm m}$ over this world-tube, it reduces to the ``particle action'' 
\begin{equation}
  \label{skeleton_action}
  S_{\rm p}=-\int m(\varphi)ds=-\int m(\varphi)
  \sqrt{g_{\mu\nu}\frac{dy_{\rm p}^\mu}{d\lambda}\frac{dy_{\rm p}^\nu}{d\lambda}}d\lambda\, ,
\end{equation}
where $ds$ is the invariant line element. 
Assume now that the stationary BH solutions of the theory~\eqref{action} are continuously connected with  the corresponding GR solution as $\alpha\rightarrow0$, and that $S_{\rm c}$ is analytic in $\varphi$.
Then the exterior metric reduces to the Kerr solution for $\alpha\to0$ and its only dimensionful parameter is the mass $M$. The corrections to the Kerr metric depend on the dimensionless parameter
\begin{equation}
  \zeta\equiv\frac{\alpha}{M^n}=q^n\frac{\alpha}{\mu^n}\,,
\end{equation}
where $q=\mu/M\ll 1$. Since $\alpha/\mu^n<1$ (otherwise GR modifications would show up in current astrophysical observations), $\zeta\ll1$, and the exterior spacetime can be approximated with the Kerr metric.

The  gravitational field equations are
\begin{equation}
  G_{\mu\nu}=R_{\mu\nu}-\frac{1}{2}g_{\mu\nu}R= T^{{\rm scal}}_{\mu\nu} + \alpha T^{c}_{\mu\nu}+ T^{\rm p}_{\mu\nu}\,,\label{eqgrav}
\end{equation}
where $T^{{\rm scal}}_{\mu\nu}=\frac{1}{2}\partial_\mu\varphi\partial_\nu\varphi-\frac{1}{4}g_{\mu\nu}(\partial \varphi)^2$ is the stress-energy tensor of the scalar field, $T^{c}_{\mu\nu}=-\frac{16\pi}{\sqrt{-g}}\frac{\delta S_c}{\delta g^{\mu\nu}}$, and
\begin{align}
  T^{{\rm p}\,\alpha\beta}
  &=8\pi
  \int  m(\varphi)\frac{\delta^{(4)}(x-y_{p}(\lambda))}{\sqrt{-g}}
  \frac{dy_p^\alpha}{d\lambda}\frac{dy_p^\beta}{d\lambda} d\lambda\,\label{def:sourceT}
\end{align}
is the stress-energy tensor of the particle. The scalar field equation is
\begin{equation}
  \square\varphi+\frac{8\pi\alpha}{\sqrt{-g}}\frac{\delta S_c}{\delta \varphi}
  =16\pi \int m'(\varphi)
        \frac{\delta^{(4)}(x-y_{p}(\lambda))}{\sqrt{-g}}d\lambda
\label{eqscalr}
\end{equation}
where $m'(\varphi)=dm(\varphi)/d\varphi$ and the $\square$ operator is evaluated on the Kerr background.

In geometric units  $[S_0]=($mass$)^2$, $[S_c]=({\rm mass})^{2-n}$. Since in the Kerr background the only dimensionful scale is the BH mass $M$ (in addition to the angular momentum $J$, which is anyway bounded by $M^2$), we expect that $S_c\sim M^{-n}S_0$. Thus, $\alpha T^{c}_{\mu\nu}\sim \zeta G_{\mu\nu}\ll G_{\mu\nu}$ and $\alpha\frac{\delta S_c}{\delta \varphi}\sim\zeta \Box\varphi\ll\Box\varphi$. The scalar field stress-energy tensor is also negligible. Indeed, in the exterior Kerr spacetime the scalar field --- due to the aforementioned no-hair theorems --- has to be a constant, $\varphi_0$, and $T^{\rm scal}_{\mu\nu}$ is quadratic in perturbations around $\varphi_0$. 

We can conclude that, under our approximations, the gravitational field equations are the same as in GR, while the scalar field equation acquires the source term on the right-hand-side of Eq.~\eqref{eqscalr}. By replacing Eq.~\eqref{sol:scalhom} we find $m'(0)/\mu=-d/4$, and since in the weak-field limit the stress-energy tensor of the particle reduces to its matter density, Eq.~\eqref{def:sourceT} leads to $m(0)=\mu$. Therefore, the scalar field equation for an EMRI in the class of theories we consider reduces to Eq.~\eqref{eq:sourcescal}.

Technically speaking, our analysis does not apply to EMRIs if the primary BH carries an appreciable scalar charge or deviates strongly from Kerr. However, it is important to emphasize that in both of these cases --~which may already be difficult to reconcile with current observations~-- one expects deviations from GR to be even larger: orbital dynamics would be further affected by the fact that the secondary object is moving in a different spacetime. Hence, our results can certainly be seen as a conservative estimate of LISA's capabilities to detect scalar fields.

%-------------------------------------------------------------
\textbf{Waveforms.} 
We study the adiabatic evolution of EMRIs  within the framework of perturbation theory, in which the secondary orbits in the Kerr background generated by the supermassive BH. Relativistic perturbations of the scalar and tensor sectors can be analysed by expanding the gravitational and scalar fields, together with the source term, in spin-weighted spheroidal harmonics\,\cite{Teukolsky:1973ha}
\begin{equation}
    \psi^{(s)} (t, r,\theta, \phi) = \int d\omega \sum_{\ell m}{ R^{(s)}_{\ell m}(r,\omega) S^{(s)}_{\ell m}(\theta,\omega) e^{i m \phi} e^{-i\omega t}},
    \label{psisdec}
\end{equation}
where tensor and scalar perturbations correspond 
to $s=-2$ and $s=0$, respectively, $(t,r\theta,\phi)$ are 
the Boyer-Lindquist coordinates, $R_{\ell m}^{(s)}$ is the 
radial part of the metric perturbations, and $S_{\ell m}^{(s)}$ 
are the spheroidal functions, specified by the mode index $\ell$ 
and by the azimuthal number $m$. 
For the scalar case  $\psi^{(s=0)} = \varphi$. The expansion yields a decoupling of the field's radial and angular  dependence, with the latter being described by the spheroidal harmonics equation. The radial component satisfies the Teukolsky equation\,\cite{Teukolsky:1973ha}
\begin{align}
\frac{1}{\Delta^s}\frac{d}{dr}\left[\Delta^{s+1}\frac{dR^{(s)}_{\ell m}}{dr}\right]+ \bigg[&\frac{K^2-2is(r-M)K}{\Delta} \nonumber\\
&+4is\omega r-\lambda\bigg]R^{(s)}_{\ell m}={\cal T}^{(s)}_{\ell m}\label{math:master}\
\end{align}
where $\Delta=r^2-2Mr+\chi^2 M^2$, $K=(r^2+\chi^2 M^2)\omega- \chi M m$, $\chi = J/M^2$, $(s,\lambda)$ are the spin-weight of the perturbed field and the spin-weighted spheroidal eigenvalue, and $\omega$ is the mode frequency. 
For the gravitational field equation, the source term ${\cal T}_{\ell m}^{(s)}$ is a combination of the components of the stress-energy tensor, expanded in spin-weighted harmonics. For the scalar field equation, assuming that the particle with scalar charge $d$ moves in equatorial circular motion,
\begin{equation}
    {\cal T}^{(s=0)}_{\ell m} = - 4 \pi \mu  d\frac{\delta(\omega-m \omega_p) \delta(r-r_p) S_{\ell m}^{* (s=0)}(\frac{\pi}{2})}{ \dot{t}}. \label{Ts0}
\end{equation}
$S_{\ell m}^{* (s=0)}(\pi/2)$ is the conjugate of the scalar spheroidal harmonic computed in $\theta=\pi/2$, $\dot{t}$ stands for the derivative of the Boyer-Lindquist coordinate $t$ with respect to the proper time, $r_p$ is the radius of the circular orbit and  
\begin{equation}
\omega_p= \frac{d\Phi}{dt}=\pm \frac{M^{1/2}}{r^{3/2}\pm \chi M^{3/2}}
\end{equation}
is the orbital frequency, where the plus (minus) sign holds for prograde (retrograde) orbits. In the following we will consider prograde orbits only. Solutions of Eq.~\eqref{math:master} can be computed by first substituting $R^{(s=0)}_{\ell m} = \psi_{\ell m}/\sqrt{r^2+\chi^2M^2}$ and computing the homogeneous solutions $\psi^{(-)}_{\ell m}$, which satisfies the boundary condition of purely ingoing wave at the horizon, and $\psi^{(+)}_{\ell m}$, which satisfies the boundary condition of purely outgoing wave at infinity. The full solutions at infinity and at the horizon are obtained integrating over the source term ${\cal T}_{\ell m}$
\begin{equation}
\psi_{\ell m}^{\pm} = \lim_{r_{\star}\rightarrow\pm\infty} e^{\pm i k^{\pm} r_{\star}} \int^{+\infty}_{-\infty} \frac{\psi^{(\mp)}_{\ell m}{\cal T}_{\ell m}}{W}\frac{\Delta}{(r^2+M^2\chi^2)^{3/2}}dr_{\star}\ ,
\end{equation}
where $k^{+} = \omega$, $k^{-} = \omega - m \Omega_{h}$,  $\omega = m\omega_p$, $\Omega_h=\chi/(2r_h)$ and $r_h=M+\sqrt{M^2-\chi^2 M^2}$. $W = \psi_{\ell m}^{'(+)}\psi_{\ell m}^{(-)}-\psi_{\ell m}^{'(-)}\psi_{\ell m}^{(+)}$ is the Wronskian with the prime denoting the derivative with respect to the tortoise coordinate $r_\star$ defined by $dr_{\star}/dr = (r^2+M^2\chi^2)/\Delta$. The full solution allows to compute the total energy flux radiated by the binary: 
\begin{equation}
\dot{E}=\sum_{i=+,-}[\dot{E}^{(i)}_{\rm grav}+\dot{E}^{(i)}_{\rm scal}]\ ,
\end{equation}
where the gravitational flux at the infinity $(+)$ and at the horizon $(-)$ can be computed as\,\cite{Hawking:1972hy,Teukolsky:1973ha,Hughes:1999bq}
\begin{equation}
\dot{E}_\tn{grv}^{(\pm)} = \sum_{\ell=2}^{\ell_{\rm max}}\sum_{m=-\ell}^{\ell}\alpha_{\ell m}^{\pm}\frac{|Z^{\pm}_{\ell m}|^2}{4 \pi \omega^2}\  \ ,
\end{equation}
where $\alpha_{\ell m}^+=1$ while $\alpha^{-}_{\ell m}$  
is given in\,\cite{Hughes:1999bq} and the scalar fluxes at the two boundaries reads\,\cite{Gralla:2005et,Warburton:2010eq}
\begin{equation}
\dot{E}_\tn{scl}^{(\pm)} = \frac{1}{16 \pi} \sum_{\ell=1}^{\ell_{\rm max}}\sum_{m=-\ell}^{\ell} \omega k^{\pm} |\delta\varphi_{\ell m}^{\pm}|^2\ . \label{math:fluxscalscalinf}
\end{equation}
The amplitudes $Z^{\pm}_{\ell m}$  can be found in the literature,\cite{Hughes:1999bq} while $\delta\varphi_{lm}^{\pm}$ are related to the scalar wave at horizon and at infinity by $ \psi_{\ell m}^{\pm}= \delta\varphi_{\ell m}^{\pm} \delta(\omega-m\omega_p)$ (the equation for the scalar flux in a previous paper\,\cite{Maselli:2020zgv} has to be corrected by replacing $1/32$ with $1/16$.)

We consider all of the multiple components in $\ddot{E}$ up to $\ell_\textnormal{max} = 10$. 
The emitted flux drives the adiabatic evolution of the inspiral
\begin{equation}
\frac{dr}{dt}=-\dot{E}\frac{dr}{dE_\tn{\rm orb}}\quad\ ,\quad \frac{d\Phi}{dt}=\omega_p\,\label{math:adiab_ev}
\end{equation}
with $E_{\rm orb}$ particle's orbital energy in the Kerr spacetime.\cite{Misner:1974qy} For a given value of the energy flux $\dot{E}$ specified by the mass ratio $q$ and by the charge $d$ we integrate Eqs.~\eqref{math:adiab_ev} with initial conditions $(r_0,\Phi_0)$ to determine the orbital evolution of the binary within a fixed observation time $T$. 

Upon the integration of Eqs.~\eqref{math:adiab_ev}, we can study the dephasing, {\it i.e.}, the difference in the GW phase evolution between an EMRI modelled with and without scalar charge.

The accumulated dephasing grows considerably during  the orbital evolution: 
for one year of inspiral before the plunge and $d=0.2$, it reaches several tens of radians after few months, and may become larger than $10^3$ in  the last month. For a signal with SNR $\sim30$ a dephasing of $0.1$~radian is considered to be resolvable.\cite{Lindblom:2008cm,Bonga:2019ycj} 
The phase difference turns out to be significantly above that threshold after twelve months of inspiral even for values of $d$ as small as $\sim5\times 10^{-3}$. 
The dephasing accumulated during the inspiral as a function of time, for different values of the scalar charge, is shown in Fig.~1 of the Supplementary material.

The inspiral trajectory obtained from Eqs.~\eqref{math:adiab_ev} allows to compute the GW signal,\cite{Barack:2003fp,Huerta:2011kt,Canizares:2012is} which we derive in the quadrupole approximation: in the transverse-traceless (TT) gauge, the metric perturbation reads
\begin{equation}
 h_{ij}^\textnormal{TT}=\frac{2}{D}\left(P_{il}P_{jm}-\frac{1}{2}P_{ij}P_{lm}\right)\ddot{I}_{lm}\ ,
\end{equation}
where $D$ is the source luminosity distance, $P_{ij}=\delta_{ij}-n_in_j$ is the projection operator onto the wave unit direction $n_j$, where $\delta_{ij}$ 
is the Kronecker delta. 
The second time derivative of the mass quadrupole moment, $\ddot{I}_{ij}$, is given 
in terms of the source stress-energy tensor
\begin{equation}
I_{ij}=\int d^3x T^{tt}(t,x^i)x^i x^j= \mu z^i(t) z^j(t)\ ,
\end{equation}
where $T^{tt}(t,x^i)=\mu\delta^{(3)}(x^i-z^i(t))$,  $x^i$ are Cartesian spatial coordinates, and $z^i(t)$ is the worldline of the secondary object in this coordinates.\cite{Babak:2006uv}
The strain produced by the GW and measured by the detector is then given by
\begin{equation}
h(t)=\frac{\sqrt{3}}{2}\left[h_+(t)F_+(t)+h_\times(t) F_\times(t)\right]\ ,\label{math:strain}
\end{equation}
where $h_+=-\left({\ddot{I}_{11}-\ddot{I}_{22}}\right)\left(1+\cos^2\!\iota\right)/2={\cal A}\cos[2\Phi(t)+2\Phi_0]\left(1+\cos^2\!\iota\right)$, $h_\times=2\ddot{I}_{12}\cos\iota=-2{\cal A}\sin[2\Phi(t)+2\Phi_0]\cos\iota$, ${\cal A}=2\mu\left[M\omega(t)\right]^{2/3}/D$ and $\iota$ is the inclination angle between the binary orbital angular momentum and the line of sight. The interferometer pattern functions $F_{\times,+}(t)$ and $\iota$ can be expressed in terms of four angles which specify the source orientation, $(\theta_\tn{s},\phi_\tn{s})$, and the direction of the BH spin, ($\theta_\tn{l},\phi_\tn{l}$) in a fixed coordinate system attached to the ecliptic.\cite{Apostolatos:1994mx,Cutler:1997ta} Their explicit expression is given in the Supplemental Material. We also correct the GW signal in order to take into account the phase modulation  induced by the LISA orbital motion.\cite{Babak:2006uv} Finally, following earlier work\,\cite{Cutler:1997ta} we assume that the LISA triangle configuration can be effectively regarded as a network of two L-shaped detectors, with the second interferometer rotated of $45^\circ$ with respect to the first one (see the Supplementary material for further details).

%-------------------------------------------------------------
\textbf{Parameter estimation.} 
We consider EMRIs in which the secondary body is moving on  equatorial circular orbits around the primary black hole. In the time domain the GW signal \eqref{math:strain} is completely determined by eleven parameters $\vec{\theta}=(\ln M,\ln \mu,\chi,\ln D,\theta_\tn{s},\phi_\tn{s},\theta_\tn{l},\phi_\tn{l},r_0,\Phi_0,d)$.
We have considered EMRIs with $M=10^6M_\odot$, $\chi=0.9$, $\mu=10M_\odot$, varying the scalar charge $d$ and fixing the source angles as $\theta_\tn{s}=\phi_\tn{s}=\pi/2$, $\theta_\tn{l}=\phi_\tn{l}=\pi/4$. We neglect the spin of the secondary body.\cite{Huerta:2011kt,Huerta:2011zi,Piovano:2021iwv} 
The initial phase has been set to $\Phi_0=0$, while the initial orbital separation is adjusted, depending on $d$, to have one year of evolution before the plunge $r_\tn{plunge}=r_\tn{ISCO}+\delta r$, where $r_\tn{ISCO}/M\simeq 2.32$ for a Kerr BH with $\chi=0.9$. 
We have fixed the radius shift to $\delta r/M=0.1$, which is more conservative than the transition region between the inspiral and the plunge as described elsewhere.\cite{Ori:2000zn} Finally, the luminosity distance $D$ is a scale factor for $h(t)$ and can be changed freely to vary the SNR of the signal (see below).

Given the waveform model we can now introduce the noise-weighted inner product between two templates
\begin{equation}
\langle h_1\vert h_2\rangle=4\Re\int_{f_\tn{min}}^{f_\tn{\rm max}}\frac{\tilde{h}_1(f)\tilde{h}^\star_2(f)}{S_n(f)}df\ , \label{math:sca_prod}
\end{equation}
where $\tilde{h}(f)$ is the Fourier transform of the time domain signal, $\tilde{h}^\star(f)$ is its complex conjugate, and $S_n(f)$ is the LISA noise spectral density.\,\cite{Cornish:2018dyw}
We sample the signal \eqref{math:strain} in the time domain, and then apply a discrete Fourier transform, evaluating the integral \eqref{math:sca_prod} between $f_\tn{min}=10^{-4}$Hz and $f_\tn{\rm max}=f_\tn{Ny}$, with $f_\tn{Ny}$ being the Nyquist frequency. The component related to the latter has been set to zero, and only Fourier components with $f\ge f_\tn{min}$ have been included. Before passing to the frequency space we taper $h(t)$ with a Tukey window with cosine fraction $\tau=0.05$. 

Eq.~\eqref{math:sca_prod} allows to determine the faithfulness 
between two waveforms:
\begin{equation}\label{eq:def_F}
\mathcal{F}[h_1,h_2]=\max_{\{t_c,\phi_c\}}\frac{\langle h_1\vert 
	h_2\rangle}{\sqrt{\langle h_1\vert h_1\rangle\langle h_2\vert h_2\rangle}}\ ,
\end{equation}
where $(t_c,\phi_c)$ are time and phase offsets.\cite{Lindblom:2008cm} 
The SNR for a specific choice of the source parameters reads $\rho=\langle h\vert h\rangle^{1/2}$. In the limit of large $\rho$ the posterior probability distribution of the source parameters, assuming flat or Gaussian priors on $\vec{\theta}$, given a certain observation $o(t)$, can be approximated by a multivariate Gaussian distribution centred around the {\it true} values,\cite{Poisson:1995ef,Vallisneri:2007ev} with covariance given by the inverse of the Fisher information matrix $\Gamma_{ij}$:
\begin{equation}
\log p(\vec{\theta}\vert o)\propto \log p_0(\theta)-\frac{1}{2}\Delta_i\Gamma_{ij}\Delta_j\ ,
\end{equation}
where $p_0$ is the parameter's prior distribution, $\Delta_i=\theta_i-\hat{\theta}_i$ is the shift between the measured and the true values $\hat{\theta}_i$, and 
\begin{equation}
\Gamma_{ij}=\left\langle \frac{\partial h}{\partial \theta_i}\bigg\vert\frac{\partial h}{\partial \theta_j}\right\rangle_{\theta=\hat{\theta}}. \label{math:def_fisher}
\end{equation}
Statistical errors on $\vec{\theta}$ and correlation coefficients among parameters are provided by diagonal and off-diagonal components of the inverse of the Fisher matrix ${\bf \Sigma}={\bf \Gamma}^{-1}$, i.e.:
\begin{equation}
\sigma_i=\Sigma^{1/2}_{ii}\quad \ ,\quad
c_{\theta_i \theta_j}=\Sigma_{ij}/(\sigma_{\theta_i}\sigma_{\theta_j})\ .
\end{equation}
Given the two-interferometer configuration for the LISA detector we can define a total SNR $\rho=\sqrt{\rho^2_\tn{I}+\rho^2_\tn{II}}$,  and a total covariance matrix of the binary parameters obtained by inverting the sum of the Fisher matrices  $\sigma^2_{\theta_i}=(\Gamma_\tn{I}+\Gamma_\tn{II})^{-1}_{ii}$.

Since waveforms are generated fully numerically in the time domain, derivatives appearing in the Fisher matrix are also numerical. We use a centered 11- and 9-point stencil for $(M,\mu,\chi,d,r_0)$ and $(\theta_\tn{s},\phi_\tn{s},\theta_\tn{l},\phi_\tn{l})$.\cite{milne2000calculus} 
For the luminosity distance $D$ and the initial phase $\Phi_0$ analytical expressions of  $\partial h/\partial\theta_i$ can be computed. Integrals are performed through Boole rule. In our $11\times 11$ parameter space, inversion of the Fisher matrices may depend on the  value of the numerical displacement chosen to compute finite difference of $h(t)$ for a specific parameter, due to numerical instability. Indeed, Fisher matrices for EMRIs are known to feature large condition numbers $\kappa={\rm max}(\lambda_i)/{\rm min}(\lambda_i)$, i.e., the ratio between the largest and the smallest eigenvalues.\cite{Gair:2012nm} 
We compute $\Gamma_{ij}$ using high-precision numerics with fluxes obtained through the BH perturbation Toolkit\,\cite{BHPToolkit} with 300 digits of input precision, which lead to  Fisher matrices of $\sim 185$ digits of final precision, and $\kappa \sim O(10^{14})$. 
Calculations of ${\bf \Gamma}$ and its inverse are extremely stable leading to discrepancies among Fisher matrices derived with different numerical shifts $\lesssim0.1\%.$ Differences on the source parameters and on the correlations coefficients are also very small, $\lesssim 1\%$ (see the Supplementary material for further details).

\section*{Acknowledgements}
\noindent
We thank Swetha Bhagwat, Jan Harms, 
Costantino Pacilio, Gabriel Piovano, Lorenzo Speri and 
Niels Warburton for useful discussions and having carefully read 
the manuscript. 
A.M. acknowledges support from the Amaldi Research Center funded 
by the MIUR program ``Dipartimento di Eccellenza'' (CUP: B81I18001170001). 
N.F. acknowledges financial support provided under the European Union's H2020 ERC Consolidator Grant ``GRavity from Astrophysical to Microscopic Scales'' grant agreement no. GRAMS-815673. 
P.P. acknowledges financial support provided under the European Union's H2020 ERC, Starting Grant agreement no.~DarkGRA--757480. 
We acknowledge financial support provided under the European Union's H2020  MSCA-RISE Grant GRU, grant agreement No. 101007855. 
T.P.S. acknowledges partial support from the STFC Consolidated Grants No. ST/T000732/1 and No. ST/V005596/1.
We also acknowledge support under the MIUR PRIN and FARE programmes (GW-NEXT, CUP:~B84I20000100001), The authors also acknowledge networking support by the COST Action GW-verse Grant No. CA16104.  All computations have been performed on the Vera cluster of the Amaldi Research Center.

%\section*{Author Contributions}
%\noindent
%A.M. and T.S. conceived the idea of probing the detectability of scalar charge by LISA with the approach developed in this manuscript. A.M. led the waveform modelling and statistical analysis and contributed to all aspects of the project. N.F. contributed significantly to the numerical calculations, the error analysis, and the interpretation of the result. L.G. and T. S. made a major contribution to the development of the theoretical framework. S.B. made an important contribution to the computation of gravitational and scalar fluxes.   L.G., T.S. and P.P. contributed strongly to the interpretation of the results and to overcoming technical difficulties in the analysis.

%\section*{Competing Interests}
%\noindent
%Authors declare no competing interests.

%\section*{Data availability statement}
%\noindent
%The data and the codes that support the plots within this paper and other findings of this study are available from the corresponding author upon request. 
%Flux calculations have been performed using numerical routines of the the Black Hole Pertubation Toolkit\,\cite{BHPToolkit}. 
%Correspondence and requests for materials should be addressed to A.M. (email: %andrea.maselli@gssi.it).

\section*{References}
\bibliographystyle{naturemag}
\bibliography{references}

\begin{thebibliography}{10}
\expandafter\ifx\csname url\endcsname\relax
  \def\url#1{\texttt{#1}}\fi
\expandafter\ifx\csname urlprefix\endcsname\relax\def\urlprefix{URL }\fi
\providecommand{\bibinfo}[2]{#2}
\providecommand{\eprint}[2][]{\url{#2}}

\bibitem{2017arXiv170200786A}
\bibinfo{author}{{Amaro-Seoane}, P.} \emph{et~al.}
\newblock \bibinfo{title}{{Laser Interferometer Space Antenna}}.
\newblock \emph{\bibinfo{journal}{arXiv e-prints}}
  \bibinfo{pages}{arXiv:1702.00786} (\bibinfo{year}{2017}).
\newblock \eprint{1702.00786}.

\bibitem{Babak:2017tow}
\bibinfo{author}{Babak, S.} \emph{et~al.}
\newblock \bibinfo{title}{{Science with the space-based interferometer LISA. V:
  Extreme mass-ratio inspirals}}.
\newblock \emph{\bibinfo{journal}{Phys. Rev. D}} \textbf{\bibinfo{volume}{95}},
  \bibinfo{pages}{103012} (\bibinfo{year}{2017}).
\newblock \eprint{1703.09722}.

\bibitem{Berti:2015itd}
\bibinfo{author}{Berti, E.} \emph{et~al.}
\newblock \bibinfo{title}{{Testing General Relativity with Present and Future
  Astrophysical Observations}}.
\newblock \emph{\bibinfo{journal}{Class. Quant. Grav.}}
  \textbf{\bibinfo{volume}{32}}, \bibinfo{pages}{243001}
  (\bibinfo{year}{2015}).
\newblock \eprint{1501.07274}.

\bibitem{Barausse:2020rsu}
\bibinfo{author}{Barausse, E.} \emph{et~al.}
\newblock \bibinfo{title}{{Prospects for Fundamental Physics with LISA}}.
\newblock \emph{\bibinfo{journal}{Gen. Rel. Grav.}}
  \textbf{\bibinfo{volume}{52}}, \bibinfo{pages}{81} (\bibinfo{year}{2020}).
\newblock \eprint{2001.09793}.

\bibitem{Copeland:2006wr}
\bibinfo{author}{Copeland, E.~J.}, \bibinfo{author}{Sami, M.} \&
  \bibinfo{author}{Tsujikawa, S.}
\newblock \bibinfo{title}{{Dynamics of dark energy}}.
\newblock \emph{\bibinfo{journal}{Int. J. Mod. Phys. D}}
  \textbf{\bibinfo{volume}{15}}, \bibinfo{pages}{1753--1936}
  (\bibinfo{year}{2006}).
\newblock \eprint{hep-th/0603057}.

\bibitem{Essig:2013lka}
\bibinfo{author}{Essig, R.} \emph{et~al.}
\newblock \bibinfo{title}{{Working Group Report: New Light Weakly Coupled
  Particles}}.
\newblock In \emph{\bibinfo{booktitle}{{Community Summer Study 2013}: {Snowmass
  on the Mississippi}}} (\bibinfo{year}{2013}).
\newblock \eprint{1311.0029}.

\bibitem{Hui:2016ltb}
\bibinfo{author}{Hui, L.}, \bibinfo{author}{Ostriker, J.~P.},
  \bibinfo{author}{Tremaine, S.} \& \bibinfo{author}{Witten, E.}
\newblock \bibinfo{title}{{Ultralight scalars as cosmological dark matter}}.
\newblock \emph{\bibinfo{journal}{Phys. Rev. D}} \textbf{\bibinfo{volume}{95}},
  \bibinfo{pages}{043541} (\bibinfo{year}{2017}).
\newblock \eprint{1610.08297}.

\bibitem{Barack:2018yly}
\bibinfo{author}{Barack, L.} \emph{et~al.}
\newblock \bibinfo{title}{{Black holes, gravitational waves and fundamental
  physics: a roadmap}}.
\newblock \emph{\bibinfo{journal}{Class. Quant. Grav.}}
  \textbf{\bibinfo{volume}{36}}, \bibinfo{pages}{143001}
  (\bibinfo{year}{2019}).
\newblock \eprint{1806.05195}.

\bibitem{Damour:1993hw}
\bibinfo{author}{Damour, T.} \& \bibinfo{author}{Esposito-Farese, G.}
\newblock \bibinfo{title}{{Nonperturbative strong field effects in tensor -
  scalar theories of gravitation}}.
\newblock \emph{\bibinfo{journal}{Phys. Rev. Lett.}}
  \textbf{\bibinfo{volume}{70}}, \bibinfo{pages}{2220--2223}
  (\bibinfo{year}{1993}).

\bibitem{Kanti:1995vq}
\bibinfo{author}{Kanti, P.}, \bibinfo{author}{Mavromatos, N.~E.},
  \bibinfo{author}{Rizos, J.}, \bibinfo{author}{Tamvakis, K.} \&
  \bibinfo{author}{Winstanley, E.}
\newblock \bibinfo{title}{{Dilatonic black holes in higher curvature string
  gravity}}.
\newblock \emph{\bibinfo{journal}{Phys. Rev. D}} \textbf{\bibinfo{volume}{54}},
  \bibinfo{pages}{5049--5058} (\bibinfo{year}{1996}).
\newblock \eprint{hep-th/9511071}.

\bibitem{Yunes:2011we}
\bibinfo{author}{Yunes, N.} \& \bibinfo{author}{Stein, L.~C.}
\newblock \bibinfo{title}{{Non-Spinning Black Holes in Alternative Theories of
  Gravity}}.
\newblock \emph{\bibinfo{journal}{Phys.\ Rev.\ D}}
  \textbf{\bibinfo{volume}{83}}, \bibinfo{pages}{104002}
  (\bibinfo{year}{2011}).
\newblock \eprint{1101.2921}.

\bibitem{Sotiriou:2013qea}
\bibinfo{author}{Sotiriou, T.~P.} \& \bibinfo{author}{Zhou, S.-Y.}
\newblock \bibinfo{title}{{Black hole hair in generalized scalar-tensor
  gravity}}.
\newblock \emph{\bibinfo{journal}{Phys. Rev. Lett.}}
  \textbf{\bibinfo{volume}{112}}, \bibinfo{pages}{251102}
  (\bibinfo{year}{2014}).
\newblock \eprint{1312.3622}.

\bibitem{zeldovich1}
\bibinfo{author}{Zel'dovich, Y.~B.}
\newblock \emph{\bibinfo{journal}{Pis'ma Zh. Eksp. Teor. Fiz.}}
  \textbf{\bibinfo{volume}{14}}, \bibinfo{pages}{270 [JETP Lett. {\bf14}, 180
  (1971)]} (\bibinfo{year}{1971}).

\bibitem{Brito:2015oca}
\bibinfo{author}{Brito, R.}, \bibinfo{author}{Cardoso, V.} \&
  \bibinfo{author}{Pani, P.}
\newblock \emph{\bibinfo{title}{{Superradiance}: {New Frontiers in Black Hole
  Physics}}}, vol. \bibinfo{volume}{971} (\bibinfo{publisher}{Springer},
  \bibinfo{year}{2020}).
\newblock \eprint{1501.06570}.

\bibitem{Arvanitaki:2009fg}
\bibinfo{author}{Arvanitaki, A.}, \bibinfo{author}{Dimopoulos, S.},
  \bibinfo{author}{Dubovsky, S.}, \bibinfo{author}{Kaloper, N.} \&
  \bibinfo{author}{March-Russell, J.}
\newblock \bibinfo{title}{{String Axiverse}}.
\newblock \emph{\bibinfo{journal}{Phys.Rev.}} \textbf{\bibinfo{volume}{D81}},
  \bibinfo{pages}{123530} (\bibinfo{year}{2010}).
\newblock \eprint{0905.4720}.

\bibitem{Cardoso:2011xi}
\bibinfo{author}{Cardoso, V.}, \bibinfo{author}{Chakrabarti, S.},
  \bibinfo{author}{Pani, P.}, \bibinfo{author}{Berti, E.} \&
  \bibinfo{author}{Gualtieri, L.}
\newblock \bibinfo{title}{{Floating and sinking: The Imprint of massive scalars
  around rotating black holes}}.
\newblock \emph{\bibinfo{journal}{Phys. Rev. Lett.}}
  \textbf{\bibinfo{volume}{107}}, \bibinfo{pages}{241101}
  (\bibinfo{year}{2011}).
\newblock \eprint{1109.6021}.

\bibitem{Silva:2017uqg}
\bibinfo{author}{Silva, H.~O.}, \bibinfo{author}{Sakstein, J.},
  \bibinfo{author}{Gualtieri, L.}, \bibinfo{author}{Sotiriou, T.~P.} \&
  \bibinfo{author}{Berti, E.}
\newblock \bibinfo{title}{{Spontaneous scalarization of black holes and compact
  stars from a Gauss-Bonnet coupling}}.
\newblock \emph{\bibinfo{journal}{Phys. Rev. Lett.}}
  \textbf{\bibinfo{volume}{120}}, \bibinfo{pages}{131104}
  (\bibinfo{year}{2018}).
\newblock \eprint{1711.02080}.

\bibitem{Doneva:2017bvd}
\bibinfo{author}{Doneva, D.~D.} \& \bibinfo{author}{Yazadjiev, S.~S.}
\newblock \bibinfo{title}{{New Gauss-Bonnet Black Holes with Curvature-Induced
  Scalarization in Extended Scalar-Tensor Theories}}.
\newblock \emph{\bibinfo{journal}{Phys. Rev. Lett.}}
  \textbf{\bibinfo{volume}{120}}, \bibinfo{pages}{131103}
  (\bibinfo{year}{2018}).
\newblock \eprint{1711.01187}.

\bibitem{Maselli:2020zgv}
\bibinfo{author}{Maselli, A.}, \bibinfo{author}{Franchini, N.},
  \bibinfo{author}{Gualtieri, L.} \& \bibinfo{author}{Sotiriou, T.~P.}
\newblock \bibinfo{title}{{Detecting scalar fields with Extreme Mass Ratio
  Inspirals}}.
\newblock \emph{\bibinfo{journal}{Phys. Rev. Lett.}}
  \textbf{\bibinfo{volume}{125}}, \bibinfo{pages}{141101}
  (\bibinfo{year}{2020}).
\newblock \eprint{2004.11895}.

\bibitem{Flanagan:1997kp}
\bibinfo{author}{Flanagan, E.~E.} \& \bibinfo{author}{Hughes, S.~A.}
\newblock \bibinfo{title}{{Measuring gravitational waves from binary black hole
  coalescences: 2. The Waves' information and its extraction, with and without
  templates}}.
\newblock \emph{\bibinfo{journal}{Phys. Rev.}} \textbf{\bibinfo{volume}{D57}},
  \bibinfo{pages}{4566--4587} (\bibinfo{year}{1998}).
\newblock \eprint{gr-qc/9710129}.

\bibitem{Lindblom:2008cm}
\bibinfo{author}{Lindblom, L.}, \bibinfo{author}{Owen, B.~J.} \&
  \bibinfo{author}{Brown, D.~A.}
\newblock \bibinfo{title}{{Model Waveform Accuracy Standards for Gravitational
  Wave Data Analysis}}.
\newblock \emph{\bibinfo{journal}{Phys. Rev.}} \textbf{\bibinfo{volume}{D78}},
  \bibinfo{pages}{124020} (\bibinfo{year}{2008}).
\newblock \eprint{0809.3844}.

\bibitem{Chatziioannou:2017tdw}
\bibinfo{author}{Chatziioannou, K.}, \bibinfo{author}{Klein, A.},
  \bibinfo{author}{Yunes, N.} \& \bibinfo{author}{Cornish, N.}
\newblock \bibinfo{title}{{Constructing Gravitational Waves from Generic
  Spin-Precessing Compact Binary Inspirals}}.
\newblock \emph{\bibinfo{journal}{Phys. Rev. D}} \textbf{\bibinfo{volume}{95}},
  \bibinfo{pages}{104004} (\bibinfo{year}{2017}).
\newblock \eprint{1703.03967}.

\bibitem{Julie:2019sab}
\bibinfo{author}{Juli\'e, F.-L.} \& \bibinfo{author}{Berti, E.}
\newblock \bibinfo{title}{{Post-Newtonian dynamics and black hole
  thermodynamics in Einstein-scalar-Gauss-Bonnet gravity}}.
\newblock \emph{\bibinfo{journal}{Phys. Rev. D}}
  \textbf{\bibinfo{volume}{100}}, \bibinfo{pages}{104061}
  (\bibinfo{year}{2019}).
\newblock \eprint{1909.05258}.

\bibitem{Perkins:2021mhb}
\bibinfo{author}{Perkins, S.~E.}, \bibinfo{author}{Nair, R.},
  \bibinfo{author}{Silva, H.~O.} \& \bibinfo{author}{Yunes, N.}
\newblock \bibinfo{title}{{Improved gravitational-wave constraints on
  higher-order curvature theories of gravity}}.
\newblock \emph{\bibinfo{journal}{Phys. Rev. D}}
  \textbf{\bibinfo{volume}{104}}, \bibinfo{pages}{024060}
  (\bibinfo{year}{2021}).
\newblock \eprint{2104.11189}.

\bibitem{Barausse:2016eii}
\bibinfo{author}{Barausse, E.}, \bibinfo{author}{Yunes, N.} \&
  \bibinfo{author}{Chamberlain, K.}
\newblock \bibinfo{title}{{Theory-Agnostic Constraints on Black-Hole Dipole
  Radiation with Multiband Gravitational-Wave Astrophysics}}.
\newblock \emph{\bibinfo{journal}{Phys. Rev. Lett.}}
  \textbf{\bibinfo{volume}{116}}, \bibinfo{pages}{241104}
  (\bibinfo{year}{2016}).
\newblock \eprint{1603.04075}.

\bibitem{Warburton:2011hp}
\bibinfo{author}{Warburton, N.} \& \bibinfo{author}{Barack, L.}
\newblock \bibinfo{title}{{Self force on a scalar charge in Kerr spacetime:
  eccentric equatorial orbits}}.
\newblock \emph{\bibinfo{journal}{Phys. Rev. D}} \textbf{\bibinfo{volume}{83}},
  \bibinfo{pages}{124038} (\bibinfo{year}{2011}).
\newblock \eprint{1103.0287}.

\bibitem{Warburton:2014bya}
\bibinfo{author}{Warburton, N.}
\newblock \bibinfo{title}{{Self force on a scalar charge in Kerr spacetime:
  inclined circular orbits}}.
\newblock \emph{\bibinfo{journal}{Phys. Rev. D}} \textbf{\bibinfo{volume}{91}},
  \bibinfo{pages}{024045} (\bibinfo{year}{2015}).
\newblock \eprint{1408.2885}.

\bibitem{Nasipak:2019hxh}
\bibinfo{author}{Nasipak, Z.}, \bibinfo{author}{Osburn, T.} \&
  \bibinfo{author}{Evans, C.~R.}
\newblock \bibinfo{title}{{Repeated faint quasinormal bursts in
  extreme-mass-ratio inspiral waveforms: Evidence from frequency-domain scalar
  self-force calculations on generic Kerr orbits}}.
\newblock \emph{\bibinfo{journal}{Phys. Rev. D}}
  \textbf{\bibinfo{volume}{100}}, \bibinfo{pages}{064008}
  (\bibinfo{year}{2019}).
\newblock \eprint{1905.13237}.

\bibitem{Cardoso:2019rou}
\bibinfo{author}{Cardoso, V.} \& \bibinfo{author}{Maselli, A.}
\newblock \bibinfo{title}{{Constraints on the astrophysical environment of
  binaries with gravitational-wave observations}}.
\newblock \emph{\bibinfo{journal}{Astron. Astrophys.}}
  \textbf{\bibinfo{volume}{644}}, \bibinfo{pages}{A147} (\bibinfo{year}{2020}).
\newblock \eprint{1909.05870}.

\bibitem{Katz:2021yft}
\bibinfo{author}{Katz, M.~L.}, \bibinfo{author}{Chua, A. J.~K.},
  \bibinfo{author}{Speri, L.}, \bibinfo{author}{Warburton, N.} \&
  \bibinfo{author}{Hughes, S.~A.}
\newblock \bibinfo{title}{{Fast extreme-mass-ratio-inspiral waveforms: New
  tools for millihertz gravitational-wave data analysis}}.
\newblock \emph{\bibinfo{journal}{Phys. Rev. D}}
  \textbf{\bibinfo{volume}{104}}, \bibinfo{pages}{064047}
  (\bibinfo{year}{2021}).
\newblock \eprint{2104.04582}.

\bibitem{Baghi:2019eqo}
\bibinfo{author}{Baghi, Q.} \emph{et~al.}
\newblock \bibinfo{title}{{Gravitational-wave parameter estimation with gaps in
  LISA: a Bayesian data augmentation method}}.
\newblock \emph{\bibinfo{journal}{Phys. Rev. D}}
  \textbf{\bibinfo{volume}{100}}, \bibinfo{pages}{022003}
  (\bibinfo{year}{2019}).
\newblock \eprint{1907.04747}.

\bibitem{Chase:1970e}
\bibinfo{author}{{Chase}, J.~E.}
\newblock \bibinfo{title}{{Event horizons in static scalar-vacuum
  space-times}}.
\newblock \emph{\bibinfo{journal}{Communications in Mathematical Physics}}
  \textbf{\bibinfo{volume}{19}}, \bibinfo{pages}{276--288}
  (\bibinfo{year}{1970}).

\bibitem{Bekenstein:1995un}
\bibinfo{author}{Bekenstein, J.~D.}
\newblock \bibinfo{title}{{Novel
  \textquoteleft{}\textquoteleft{}no-scalar-hair\textquoteright{}\textquoteright{}
  theorem for black holes}}.
\newblock \emph{\bibinfo{journal}{Phys. Rev. D}} \textbf{\bibinfo{volume}{51}},
  \bibinfo{pages}{R6608} (\bibinfo{year}{1995}).

\bibitem{Hawking:1972qk}
\bibinfo{author}{Hawking, S.~W.}
\newblock \bibinfo{title}{{Black holes in the Brans-Dicke theory of
  gravitation}}.
\newblock \emph{\bibinfo{journal}{Commun. Math. Phys.}}
  \textbf{\bibinfo{volume}{25}}, \bibinfo{pages}{167--171}
  (\bibinfo{year}{1972}).

\bibitem{Sotiriou:2011dz}
\bibinfo{author}{Sotiriou, T.~P.} \& \bibinfo{author}{Faraoni, V.}
\newblock \bibinfo{title}{{Black holes in scalar-tensor gravity}}.
\newblock \emph{\bibinfo{journal}{Phys. Rev. Lett.}}
  \textbf{\bibinfo{volume}{108}}, \bibinfo{pages}{081103}
  (\bibinfo{year}{2012}).
\newblock \eprint{1109.6324}.

\bibitem{Hui:2012qt}
\bibinfo{author}{Hui, L.} \& \bibinfo{author}{Nicolis, A.}
\newblock \bibinfo{title}{{No-Hair Theorem for the Galileon}}.
\newblock \emph{\bibinfo{journal}{Phys. Rev. Lett.}}
  \textbf{\bibinfo{volume}{110}}, \bibinfo{pages}{241104}
  (\bibinfo{year}{2013}).
\newblock \eprint{1202.1296}.

\bibitem{Cardoso:2016ryw}
\bibinfo{author}{Cardoso, V.} \& \bibinfo{author}{Gualtieri, L.}
\newblock \bibinfo{title}{{Testing the black hole
  \textquoteleft{}no-hair\textquoteright{} hypothesis}}.
\newblock \emph{\bibinfo{journal}{Class. Quant. Grav.}}
  \textbf{\bibinfo{volume}{33}}, \bibinfo{pages}{174001}
  (\bibinfo{year}{2016}).
\newblock \eprint{1607.03133}.

\bibitem{Sotiriou:2014pfa}
\bibinfo{author}{Sotiriou, T.~P.} \& \bibinfo{author}{Zhou, S.-Y.}
\newblock \bibinfo{title}{{Black hole hair in generalized scalar-tensor
  gravity: An explicit example}}.
\newblock \emph{\bibinfo{journal}{Phys. Rev. D}} \textbf{\bibinfo{volume}{90}},
  \bibinfo{pages}{124063} (\bibinfo{year}{2014}).
\newblock \eprint{1408.1698}.

\bibitem{Yunes:2011aa}
\bibinfo{author}{Yunes, N.}, \bibinfo{author}{Pani, P.} \&
  \bibinfo{author}{Cardoso, V.}
\newblock \bibinfo{title}{{Gravitational Waves from Quasicircular Extreme
  Mass-Ratio Inspirals as Probes of Scalar-Tensor Theories}}.
\newblock \emph{\bibinfo{journal}{Phys. Rev. D}} \textbf{\bibinfo{volume}{85}},
  \bibinfo{pages}{102003} (\bibinfo{year}{2012}).
\newblock \eprint{1112.3351}.

\bibitem{Hannuksela:2018izj}
\bibinfo{author}{Hannuksela, O.~A.}, \bibinfo{author}{Wong, K. W.~K.},
  \bibinfo{author}{Brito, R.}, \bibinfo{author}{Berti, E.} \&
  \bibinfo{author}{Li, T. G.~F.}
\newblock \bibinfo{title}{{Probing the existence of ultralight bosons with a
  single gravitational-wave measurement}}.
\newblock \emph{\bibinfo{journal}{Nature Astron.}}
  \textbf{\bibinfo{volume}{3}}, \bibinfo{pages}{447--451}
  (\bibinfo{year}{2019}).
\newblock \eprint{1804.09659}.

\bibitem{Annulli:2020ilw}
\bibinfo{author}{Annulli, L.}, \bibinfo{author}{Cardoso, V.} \&
  \bibinfo{author}{Vicente, R.}
\newblock \bibinfo{title}{{Stirred and shaken: Dynamical behavior of boson
  stars and dark matter cores}}.
\newblock \emph{\bibinfo{journal}{Phys. Lett. B}}
  \textbf{\bibinfo{volume}{811}}, \bibinfo{pages}{135944}
  (\bibinfo{year}{2020}).
\newblock \eprint{2007.03700}.

\bibitem{Eardley:1975sc}
\bibinfo{author}{{Eardley}, D.~M.}
\newblock \bibinfo{title}{{Observable effects of a scalar gravitational field
  in a binary pulsar.}}
\newblock \emph{\bibinfo{journal}{The Astrophysical Journal Letters}}
  \textbf{\bibinfo{volume}{196}}, \bibinfo{pages}{L59--L62}
  (\bibinfo{year}{1975}).

\bibitem{Damour:1992we}
\bibinfo{author}{Damour, T.} \& \bibinfo{author}{Esposito-Farese, G.}
\newblock \bibinfo{title}{{Tensor multiscalar theories of gravitation}}.
\newblock \emph{\bibinfo{journal}{Class. Quant. Grav.}}
  \textbf{\bibinfo{volume}{9}}, \bibinfo{pages}{2093--2176}
  (\bibinfo{year}{1992}).

\bibitem{Julie:2017ucp}
\bibinfo{author}{Juli\'e, F.-L.}
\newblock \bibinfo{title}{{Reducing the two-body problem in scalar-tensor
  theories to the motion of a test particle : a scalar-tensor
  effective-one-body approach}}.
\newblock \emph{\bibinfo{journal}{Phys. Rev. D}} \textbf{\bibinfo{volume}{97}},
  \bibinfo{pages}{024047} (\bibinfo{year}{2018}).
\newblock \eprint{1709.09742}.

\bibitem{Julie:2017rpw}
\bibinfo{author}{Juli\'e, F.-L.}
\newblock \bibinfo{title}{{On the motion of hairy black holes in
  Einstein-Maxwell-dilaton theories}}.
\newblock \emph{\bibinfo{journal}{JCAP}} \textbf{\bibinfo{volume}{01}},
  \bibinfo{pages}{026} (\bibinfo{year}{2018}).
\newblock \eprint{1711.10769}.

\bibitem{Steinhoff:2012rw}
\bibinfo{author}{Steinhoff, J.} \& \bibinfo{author}{Puetzfeld, D.}
\newblock \bibinfo{title}{{Influence of internal structure on the motion of
  test bodies in extreme mass ratio situations}}.
\newblock \emph{\bibinfo{journal}{Phys. Rev. D}} \textbf{\bibinfo{volume}{86}},
  \bibinfo{pages}{044033} (\bibinfo{year}{2012}).
\newblock \eprint{1205.3926}.

\bibitem{Teukolsky:1973ha}
\bibinfo{author}{Teukolsky, S.~A.}
\newblock \bibinfo{title}{{Perturbations of a rotating black hole. 1.
  Fundamental equations for gravitational electromagnetic and neutrino field
  perturbations}}.
\newblock \emph{\bibinfo{journal}{Astrophys. J.}}
  \textbf{\bibinfo{volume}{185}}, \bibinfo{pages}{635--647}
  (\bibinfo{year}{1973}).

\bibitem{Hawking:1972hy}
\bibinfo{author}{Hawking, S.~W.} \& \bibinfo{author}{Hartle, J.~B.}
\newblock \bibinfo{title}{{Energy and angular momentum flow into a black
  hole}}.
\newblock \emph{\bibinfo{journal}{Commun. Math. Phys.}}
  \textbf{\bibinfo{volume}{27}}, \bibinfo{pages}{283--290}
  (\bibinfo{year}{1972}).

\bibitem{Hughes:1999bq}
\bibinfo{author}{Hughes, S.~A.}
\newblock \bibinfo{title}{{The Evolution of circular, nonequatorial orbits of
  Kerr black holes due to gravitational wave emission}}.
\newblock \emph{\bibinfo{journal}{Phys. Rev. D}} \textbf{\bibinfo{volume}{61}},
  \bibinfo{pages}{084004} (\bibinfo{year}{2000}).
\newblock \bibinfo{note}{[Erratum: Phys.Rev.D 63, 049902 (2001), Erratum:
  Phys.Rev.D 65, 069902 (2002), Erratum: Phys.Rev.D 67, 089901 (2003), Erratum:
  Phys.Rev.D 78, 109902 (2008), Erratum: Phys.Rev.D 90, 109904 (2014)]},
  \eprint{gr-qc/9910091}.

\bibitem{Gralla:2005et}
\bibinfo{author}{Gralla, S.~E.}, \bibinfo{author}{Friedman, J.~L.} \&
  \bibinfo{author}{Wiseman, A.~G.}
\newblock \bibinfo{title}{{Numerical radiation reaction for a scalar charge in
  Kerr circular orbit}}  (\bibinfo{year}{2005}).
\newblock \eprint{gr-qc/0502123}.

\bibitem{Warburton:2010eq}
\bibinfo{author}{Warburton, N.} \& \bibinfo{author}{Barack, L.}
\newblock \bibinfo{title}{{Self force on a scalar charge in Kerr spacetime:
  circular equatorial orbits}}.
\newblock \emph{\bibinfo{journal}{Phys. Rev. D}} \textbf{\bibinfo{volume}{81}},
  \bibinfo{pages}{084039} (\bibinfo{year}{2010}).
\newblock \eprint{1003.1860}.

\bibitem{Misner:1974qy}
\bibinfo{author}{Misner, C.~W.}, \bibinfo{author}{Thorne, K.~S.} \&
  \bibinfo{author}{Wheeler, J.~A.}
\newblock \emph{\bibinfo{title}{{Gravitation}}} (\bibinfo{publisher}{W. H.
  Freeman}, \bibinfo{address}{San Francisco}, \bibinfo{year}{1973}).

\bibitem{Bonga:2019ycj}
\bibinfo{author}{Bonga, B.}, \bibinfo{author}{Yang, H.} \&
  \bibinfo{author}{Hughes, S.~A.}
\newblock \bibinfo{title}{{Tidal resonance in extreme mass-ratio inspirals}}.
\newblock \emph{\bibinfo{journal}{Phys. Rev. Lett.}}
  \textbf{\bibinfo{volume}{123}}, \bibinfo{pages}{101103}
  (\bibinfo{year}{2019}).
\newblock \eprint{1905.00030}.

\bibitem{Barack:2003fp}
\bibinfo{author}{Barack, L.} \& \bibinfo{author}{Cutler, C.}
\newblock \bibinfo{title}{{LISA capture sources: Approximate waveforms,
  signal-to-noise ratios, and parameter estimation accuracy}}.
\newblock \emph{\bibinfo{journal}{Phys. Rev. D}} \textbf{\bibinfo{volume}{69}},
  \bibinfo{pages}{082005} (\bibinfo{year}{2004}).
\newblock \eprint{gr-qc/0310125}.

\bibitem{Huerta:2011kt}
\bibinfo{author}{Huerta, E.~A.} \& \bibinfo{author}{Gair, J.~R.}
\newblock \bibinfo{title}{{Importance of including small body spin effects in
  the modelling of extreme and intermediate mass-ratio inspirals}}.
\newblock \emph{\bibinfo{journal}{Phys. Rev. D}} \textbf{\bibinfo{volume}{84}},
  \bibinfo{pages}{064023} (\bibinfo{year}{2011}).
\newblock \eprint{1105.3567}.

\bibitem{Canizares:2012is}
\bibinfo{author}{Canizares, P.}, \bibinfo{author}{Gair, J.~R.} \&
  \bibinfo{author}{Sopuerta, C.~F.}
\newblock \bibinfo{title}{{Testing Chern-Simons Modified Gravity with
  Gravitational-Wave Detections of Extreme-Mass-Ratio Binaries}}.
\newblock \emph{\bibinfo{journal}{Phys. Rev. D}} \textbf{\bibinfo{volume}{86}},
  \bibinfo{pages}{044010} (\bibinfo{year}{2012}).
\newblock \eprint{1205.1253}.

\bibitem{Babak:2006uv}
\bibinfo{author}{Babak, S.}, \bibinfo{author}{Fang, H.}, \bibinfo{author}{Gair,
  J.~R.}, \bibinfo{author}{Glampedakis, K.} \& \bibinfo{author}{Hughes, S.~A.}
\newblock \bibinfo{title}{{'Kludge' gravitational waveforms for a test-body
  orbiting a Kerr black hole}}.
\newblock \emph{\bibinfo{journal}{Phys. Rev. D}} \textbf{\bibinfo{volume}{75}},
  \bibinfo{pages}{024005} (\bibinfo{year}{2007}).
\newblock \bibinfo{note}{[Erratum: Phys.Rev.D 77, 04990 (2008)]},
  \eprint{gr-qc/0607007}.

\bibitem{Apostolatos:1994mx}
\bibinfo{author}{Apostolatos, T.~A.}, \bibinfo{author}{Cutler, C.},
  \bibinfo{author}{Sussman, G.~J.} \& \bibinfo{author}{Thorne, K.~S.}
\newblock \bibinfo{title}{{Spin induced orbital precession and its modulation
  of the gravitational wave forms from merging binaries}}.
\newblock \emph{\bibinfo{journal}{Phys. Rev. D}} \textbf{\bibinfo{volume}{49}},
  \bibinfo{pages}{6274--6297} (\bibinfo{year}{1994}).

\bibitem{Cutler:1997ta}
\bibinfo{author}{Cutler, C.}
\newblock \bibinfo{title}{{Angular resolution of the LISA gravitational wave
  detector}}.
\newblock \emph{\bibinfo{journal}{Phys. Rev. D}} \textbf{\bibinfo{volume}{57}},
  \bibinfo{pages}{7089--7102} (\bibinfo{year}{1998}).
\newblock \eprint{gr-qc/9703068}.

\bibitem{Huerta:2011zi}
\bibinfo{author}{Huerta, E.~A.}, \bibinfo{author}{Gair, J.~R.} \&
  \bibinfo{author}{Brown, D.~A.}
\newblock \bibinfo{title}{{Importance of including small body spin effects in
  the modelling of intermediate mass-ratio inspirals. II Accurate parameter
  extraction of strong sources using higher-order spin effects}}.
\newblock \emph{\bibinfo{journal}{Phys. Rev. D}} \textbf{\bibinfo{volume}{85}},
  \bibinfo{pages}{064023} (\bibinfo{year}{2012}).
\newblock \eprint{1111.3243}.

\bibitem{Piovano:2021iwv}
\bibinfo{author}{Piovano, G.~A.}, \bibinfo{author}{Brito, R.},
  \bibinfo{author}{Maselli, A.} \& \bibinfo{author}{Pani, P.}
\newblock \bibinfo{title}{{Assessing the detectability of the secondary spin in
  extreme mass-ratio inspirals with fully relativistic numerical waveforms}}.
\newblock \emph{\bibinfo{journal}{Phys. Rev. D}}
  \textbf{\bibinfo{volume}{104}}, \bibinfo{pages}{124019}
  (\bibinfo{year}{2021}).
\newblock \eprint{2105.07083}.

\bibitem{Ori:2000zn}
\bibinfo{author}{Ori, A.} \& \bibinfo{author}{Thorne, K.~S.}
\newblock \bibinfo{title}{{The Transition from inspiral to plunge for a compact
  body in a circular equatorial orbit around a massive, spinning black hole}}.
\newblock \emph{\bibinfo{journal}{Phys. Rev. D}} \textbf{\bibinfo{volume}{62}},
  \bibinfo{pages}{124022} (\bibinfo{year}{2000}).
\newblock \eprint{gr-qc/0003032}.

\bibitem{Cornish:2018dyw}
\bibinfo{author}{Robson, T.}, \bibinfo{author}{Cornish, N.~J.} \&
  \bibinfo{author}{Liu, C.}
\newblock \bibinfo{title}{{The construction and use of LISA sensitivity
  curves}}.
\newblock \emph{\bibinfo{journal}{Class. Quant. Grav.}}
  \textbf{\bibinfo{volume}{36}}, \bibinfo{pages}{105011}
  (\bibinfo{year}{2019}).
\newblock \eprint{1803.01944}.

\bibitem{Poisson:1995ef}
\bibinfo{author}{Poisson, E.} \& \bibinfo{author}{Will, C.~M.}
\newblock \bibinfo{title}{{Gravitational waves from inspiraling compact
  binaries: Parameter estimation using second postNewtonian wave forms}}.
\newblock \emph{\bibinfo{journal}{Phys. Rev. D}} \textbf{\bibinfo{volume}{52}},
  \bibinfo{pages}{848--855} (\bibinfo{year}{1995}).
\newblock \eprint{gr-qc/9502040}.

\bibitem{Vallisneri:2007ev}
\bibinfo{author}{Vallisneri, M.}
\newblock \bibinfo{title}{{Use and abuse of the Fisher information matrix in
  the assessment of gravitational-wave parameter-estimation prospects}}.
\newblock \emph{\bibinfo{journal}{Phys. Rev. D}} \textbf{\bibinfo{volume}{77}},
  \bibinfo{pages}{042001} (\bibinfo{year}{2008}).
\newblock \eprint{gr-qc/0703086}.

\bibitem{milne2000calculus}
\bibinfo{author}{Milne-Thomson, L.~M.}
\newblock \emph{\bibinfo{title}{The Calculus of Finite Differences}},
  vol.~\bibinfo{volume}{15} (\bibinfo{publisher}{Cambridge University Press},
  \bibinfo{year}{1936}).

\bibitem{Gair:2012nm}
\bibinfo{author}{Gair, J.~R.}, \bibinfo{author}{Vallisneri, M.},
  \bibinfo{author}{Larson, S.~L.} \& \bibinfo{author}{Baker, J.~G.}
\newblock \bibinfo{title}{{Testing General Relativity with Low-Frequency,
  Space-Based Gravitational-Wave Detectors}}.
\newblock \emph{\bibinfo{journal}{Living Rev. Rel.}}
  \textbf{\bibinfo{volume}{16}}, \bibinfo{pages}{7} (\bibinfo{year}{2013}).
\newblock \eprint{1212.5575}.

\bibitem{BHPToolkit}
\bibinfo{title}{{Black Hole Perturbation Toolkit}}.
\newblock
  \bibinfo{howpublished}{(\href{http://bhptoolkit.org/}{bhptoolkit.org})}.

\bibitem{Mathematica}
\bibinfo{author}{Inc., W.~R.}
\newblock \bibinfo{title}{Mathematica, {V}ersion 12.3.1}.
\newblock \urlprefix\url{https://www.wolfram.com/mathematica}.
\newblock \bibinfo{note}{Champaign, IL, 2021}.

\bibitem{Milne-ThomsonLouisMelville1951Tcof}
\bibinfo{author}{Milne-Thomson, L.~M.}
\newblock \emph{\bibinfo{title}{The calculus of finite differences / by L. M.
  Milne-Thomson}} (\bibinfo{publisher}{Macmillan and Co},
  \bibinfo{address}{London}, \bibinfo{year}{1951}).

\bibitem{Press:1992zz}
\bibinfo{author}{Press, W.~H.}, \bibinfo{author}{Teukolsky, S.~A.},
  \bibinfo{author}{Vetterling, W.~T.} \& \bibinfo{author}{Flannery, B.~P.}
\newblock \bibinfo{title}{{Numerical Recipes in FORTRAN: The Art of Scientific
  Computing}}  (\bibinfo{year}{1992}).

\bibitem{Porter:2009wi}
\bibinfo{author}{Porter, E.~K.}
\newblock \bibinfo{title}{{An Overview of LISA Data Analysis Algorithms}}
  (\bibinfo{year}{2009}).
\newblock \eprint{0910.0373}.

\bibitem{Porter:2015eha}
\bibinfo{author}{Porter, E.~K.} \& \bibinfo{author}{Cornish, N.~J.}
\newblock \bibinfo{title}{{Fisher versus Bayes: A comparison of parameter
  estimation techniques for massive black hole binaries to high redshifts with
  eLISA}}.
\newblock \emph{\bibinfo{journal}{Phys. Rev. D}} \textbf{\bibinfo{volume}{91}},
  \bibinfo{pages}{104001} (\bibinfo{year}{2015}).
\newblock \eprint{1502.05735}.

\bibitem{Speri:2021psr}
\bibinfo{author}{Speri, L.} \& \bibinfo{author}{Gair, J.~R.}
\newblock \bibinfo{title}{{Assessing the impact of transient orbital
  resonances}}  (\bibinfo{year}{2021}).
\newblock \eprint{2103.06306}.

\bibitem{Pai:2012mv}
\bibinfo{author}{Pai, A.} \& \bibinfo{author}{Arun, K.~G.}
\newblock \bibinfo{title}{{Singular value decomposition in parametrised tests
  of post-Newtonian theory}}.
\newblock \emph{\bibinfo{journal}{Class. Quant. Grav.}}
  \textbf{\bibinfo{volume}{30}}, \bibinfo{pages}{025011}
  (\bibinfo{year}{2013}).
\newblock \eprint{1207.1943}.

\bibitem{Wade:2013hoa}
\bibinfo{author}{Wade, M.}, \bibinfo{author}{Creighton, J. D.~E.},
  \bibinfo{author}{Ochsner, E.} \& \bibinfo{author}{Nielsen, A.~B.}
\newblock \bibinfo{title}{{Advanced LIGO\textquoteright{}s ability to detect
  apparent violations of the cosmic censorship conjecture and the no-hair
  theorem through compact binary coalescence detections}}.
\newblock \emph{\bibinfo{journal}{Phys. Rev. D}} \textbf{\bibinfo{volume}{88}},
  \bibinfo{pages}{083002} (\bibinfo{year}{2013}).
\newblock \eprint{1306.3901}.

\end{thebibliography}
\noindent

%-------------------------------------------------------------
%-------------------------------------------------------------
\section*{Supplementary material for: Detecting new 
	fundamental fields with LISA}
\noindent

%-------------------------------------------------------------
\textbf{LISA configuration.} 
We consider the LISA spectral density (PSD) of Cornish \& Robson\,\cite{Cornish:2018dyw} who provide an accurate analytic fit for the detector noise. 
The PSD consists of two parts: the instrumental and the confusion noise produced by unresolved galactic binaries, i.e. 
\begin{equation}
	S_{n}(f)=S^\tn{Ins}_{n}(f)+S^\tn{WDN}_{n}(f) \ .
\end{equation}
where 
\begin{equation}
	S^\tn{Ins}_{n}(f)=A_1\left(P_\tn{OMS}+2[1+\cos^2(f/f_\star)]
	\frac{P_\tn{acc}}{(2\pi f)^4}\right)
	\left(1+\frac{6}{10}\frac{f^2}{f_\star^2}\right)\ ,\nonumber
\end{equation}
$A_1=\frac{10}{3L^2}$, $L=2.5$Gm, $f_\star=19.09$mHz, while 
\begin{align}
	P_\tn{OMS}&=(1.5\times 10^{-11}\tn{m}^2\left[1+\left(\frac{2\tn{mHz}}{f}\right)^4\right]\ \tn{Hz}^{-1}\nonumber\ ,\\ 
	P_\tn{ACC}&=(3\times 10^{-15}\tn{ms}^{-2})^2\left[1+\left(\frac{0.4\tn{mHz}}{f}\right)^2\right]\times \nonumber \\
	&\phantom{aaaaaaaaaaaaaaaaaaaaaaaaaaaaa}\left[1+\left(\frac{f}{8\tn{mHz}}\right)^4\right]\ \tn{Hz}^{-1}\nonumber\ .
\end{align}
For the white dwarf contribution
\begin{equation}
	S^\tn{WDN}_{n}=A_2f^{-7/3}e^{-f^\alpha+\beta f\sin(\kappa f)}[1+\tanh(\gamma(f_k-f))]\ \tn{Hz}^{-1}\ , \nonumber
\end{equation}
with the amplitude $A_2=9\times 10^{-45}$, and the coefficients 
$(\alpha,\beta,\kappa,\gamma,f_k)=(0.171,292,1020,1680,0.00215)$.

%
%%%%%%%%%%%%%%%%%%%%%%%%%%%%%%%%%%%%%%%%%%%%%%%%%%%%%%%%%%%%%%%%
\begin{figure}
	\includegraphics[width=\columnwidth]{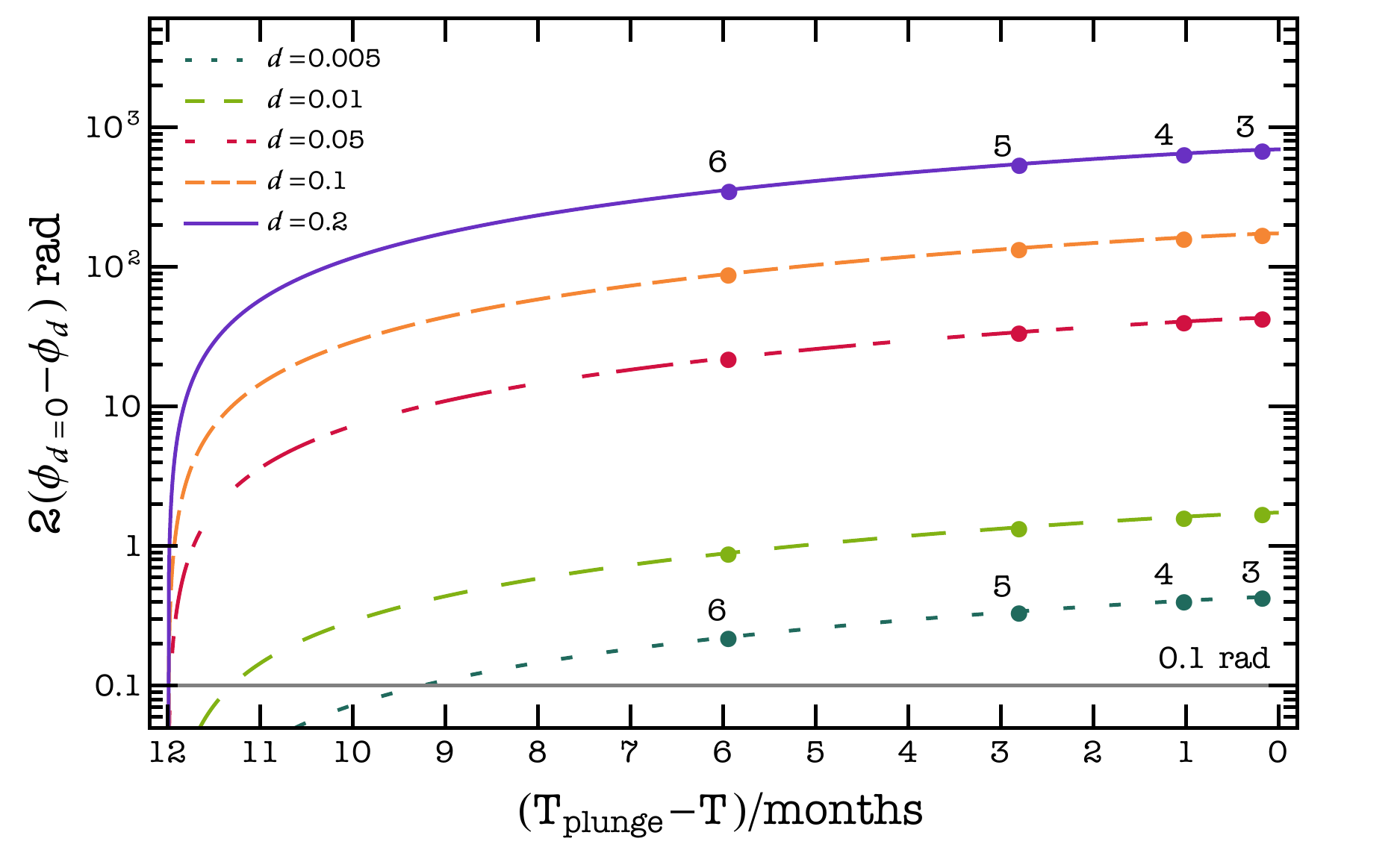}
	\caption{
		\textbf{Difference in the GW phase evolution of EMRIs with and without scalar charge.} 
		Colored curves correspond to  different values of $d$. Numbered bullet  points identify the  values of the orbital radius $r/M$ crossed by the corresponding binary system with $d\neq 0$ during the inspiral. The horizontal  line identifies the threshold for phase  resolution by LISA,  assuming a signal with SNR of $30$.} \label{fig:dephasing}
\end{figure}
%%%%%%%%%%%%%%%%%%%%%%%%%%%%%%%%%%%%%%%%%%%%%%%%%%%%%%%%%%%%%%%%
%

\begin{figure}
	\centering
	\includegraphics[width=\columnwidth]{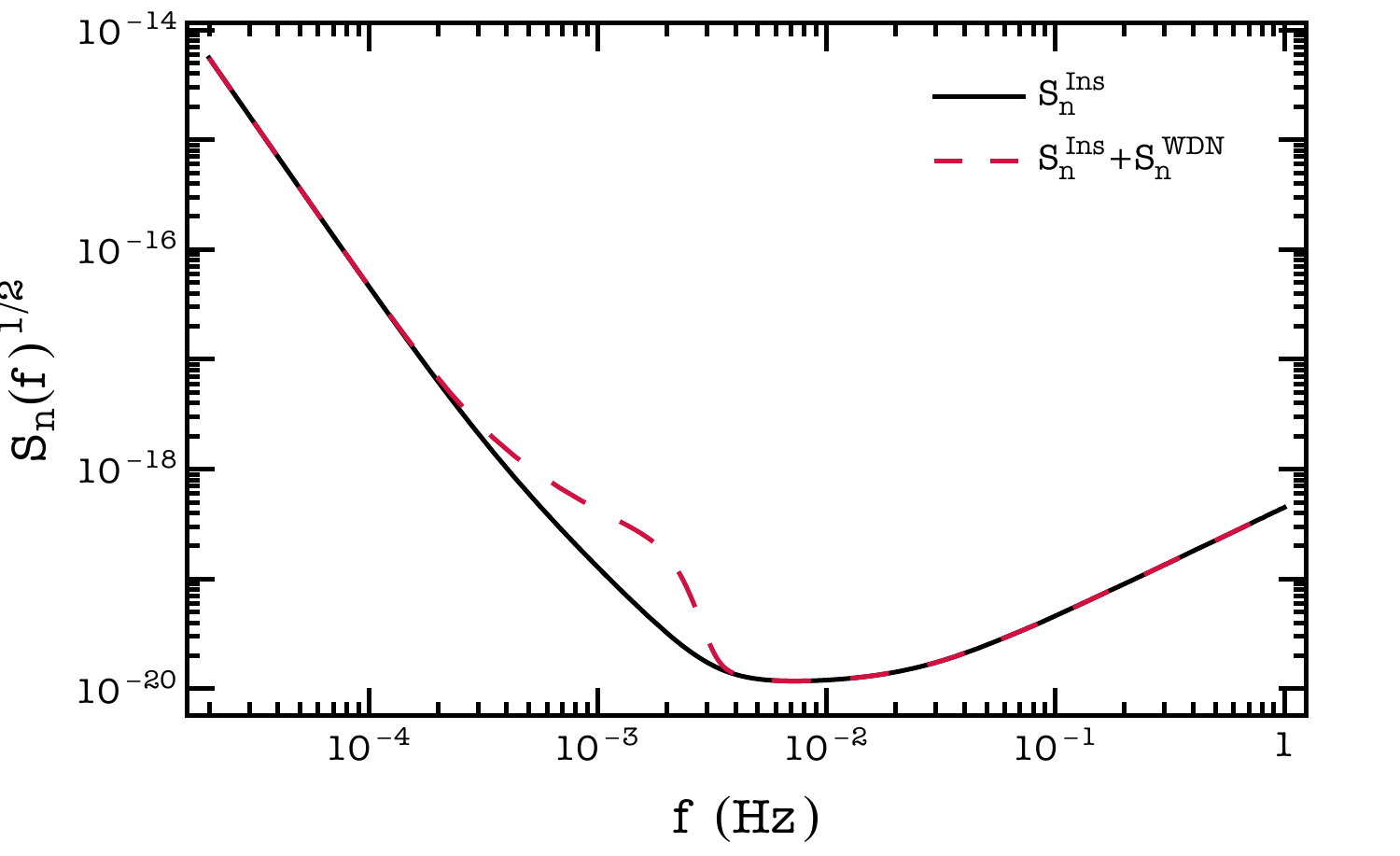}
	\caption{
		\textbf{Noise spectral density for LISA as a function of the 
			frequency, with and without the confusion noise produced by 
			unresolved galactic white dwarf binaries (WDN).}}
	\label{fig:LISAPSD}
\end{figure}

The PSD is shown in Figure~\ref{fig:LISAPSD}.
The gravitational wave strain measured by LISA is given by: 
\begin{equation}
	h(t)=\frac{\sqrt{3}}{2} \left[h_+(t)F_+(\theta,\phi,\psi)+h_\times(t) F_\times(\theta,\phi,\psi) \right]\ ,\label{math:strain2}
\end{equation}
where the interferometer's pattern functions $F_{\times,+}$ are defined as:
\begin{align}
	F_+=\frac{1+\cos^2\theta}{2}&\cos2\phi\cos2\psi-\cos\theta\sin2\phi\sin2\psi\ ,\nonumber\\
	F_\times=\frac{1+\cos^2\theta}{2}&\cos2\phi\sin2\psi+\cos\theta\sin2\phi\cos2\psi\ .\nonumber
\end{align}
Here $(\theta,\phi)$ describe the location of the EMRI in the sky and $\psi$ is the polarization angle,\cite{Apostolatos:1994mx} all defined in the detector reference frame. Given LISA orbital motion around the Sun they also depend on time, and it is more convenient to recast these angles in a fixed coordinate system attached to the ecliptic.\cite{Huerta:2011kt} In this set-up
\begin{align}
	\cos\theta(t)&=\frac{1}{2}\cos\theta_\tn{s}-\frac{\sqrt{3}}{2}\sin\theta_\tn{s}\cos[\phi_t-\phi_\tn{s}]\ ,\nonumber\\
	\phi(t)&=\alpha_0+\phi_t+\tan^{-1}\left[\frac{\sqrt{3}\cos\theta_\tn{s}+\sin{\theta}_\tn{s}\cos[\phi_t-
		\phi_\tn{s}]}{2\sin\theta_\tn{s}\sin[\phi_t-\phi_\tn{s}]}\right]\ ,
\end{align}
where $\phi_t=\beta_0+2\pi t/T$, $T=1$ year is the LISA orbital period, and $(\theta_s,\phi_s)$ are constants that identify the binary orientation with respect to the fixed reference frame. 
The two angles $(\beta_0,\alpha_0)$ specify the orbital and rotational phase of the interferometer at $t=0$, and can be set both equal to zero. The polarization angle as a function of time can be written in a compact form as 
\begin{equation}
	\psi(t)=\tan^{-1}\frac{\hat{L}\cdot\hat{z}-(\hat{L}\cdot\hat{N})(\hat{z}\cdot \hat{N})}{\hat{N}\cdot(\hat{L}\times\hat{z})}\ ,
\end{equation}
with $\hat{z}\cdot\hat{N}=\cos\theta(t)$ and 
\begin{align}
	\cos\iota=\hat{L}\cdot\hat{N}&=\cos\theta_\tn{l}\cos\theta_\tn{s}+\sin\theta_\tn{l}\sin\theta_\tn{s}
	\cos[\phi_\tn{l}-\phi_\tn{s}]\ ,\nonumber\\
	\hat{L}\cdot\hat{z}&=\frac{1}{2}\cos\theta_\tn{l}-\frac{\sqrt{3}}{2}\sin\theta_\tn{l}\cos[\phi_t-\phi_\tn{l}]\ ,\nonumber\\
	\hat{N}\cdot(\hat{L}\times\hat{z})&=\frac{1}{2}\sin\theta_\tn{l}\sin\theta_\tn{s}
	\sin[\phi_\tn{l}-\phi_\tn{s}]\nonumber\\
	&-\frac{\sqrt{3}}{2}\cos\phi_t\left[\cos\theta_\tn{l}\sin\theta_\tn{s}\sin\phi_\tn{s}-\cos\theta_\tn{s}\sin\theta_\tn{l}\sin\phi_\tn{l}\right]\nonumber\\
	&-\frac{\sqrt{3}}{2}\sin\phi_t\left[\cos\theta_\tn{s}\sin\theta_\tn{l}\cos\phi_\tn{l}-\cos\theta_\tn{l}\sin\theta_\tn{s}\cos\phi_\tn{s}\right]\ ,
\end{align}
where $(\theta_\tn{l},\phi_\tn{l})$ define the direction of the binary's angular momentum in the ecliptic reference system. Moreover, as discussed by Cutler,\cite{Cutler:1997ta} the LISA triangle configuration can be effectively regarded as a network of two L-shaped detectors, with the second interferometer rotated of $45^\circ$ with respect to the first one. We can therefore define two sets of pattern functions, $F^\tn{I,II}_{\times,+}$, which correspond to each of the two detectors and such that $F^\tn{II}_{\times,+}=F^\tn{I}_{\times,+}(\theta,\phi-\pi/4,\psi)$.

Finally, the gravitational wave signal also acquires a modulation due to the LISA orbital motion.\cite{Babak:2006uv} We correct for this effect by modifying the phase of the waveform as 
\begin{equation}
	\Phi(t)\rightarrow \Phi(t)+\Phi'(t) R_\tn{AU} 
	\sin\theta_s \cos(2\pi t/T-\phi_s)\ ,
\end{equation}
where $R_\tn{AU}$ is the astronomical unit and $T=1$ year.

%-------------------------------------------------------------
\textbf{Numerical setup.} 
In this section we provide technical details as well as a more exhaustive discussion on the features and on the accuracy of the numerical calculations we performed. The examples we provide refer to binary configurations with the source parameters considered in the main text, and in particular for binaries with $M=10^6M_\odot$, $\chi=0.9$ and $\mu=10M_\odot$, and SNR 150.

The sensitivity to small changes during the inspiral, along with the long and computationally expensive waveforms, render parameter estimation of EMRIs a challenging task that requires high precision methods.\cite{Gair:2012nm} We make use of the Black Hole Perturbation Toolkit\,\cite{BHPToolkit} for the computation of the tensor and scalar energy fluxes. In both cases we have summed multipole contributions up to $\ell=10$. Fluxes as a function of the orbital radius have been sampled on a grid of 100 equally spaced points within $r\in [8,2.4]M$, with  300 digits of input precision. In order to perform spin derivatives we have computed $\dot{E}$ sampling 11 points of $\chi$ in uniform steps of $\Delta \chi= 0.02$ symmetrically around $\chi=0.9$. Given that the dependence on the scalar charge can be factored out analytically from the fluxes, the overall gravitational wave luminosity which sources the EMRI phase evolution, needs to be numerically interpolated only along the radial coordinate and the spin, namely
\begin{equation}   
	\dot{E}(r,a,d)=\dot{E}_\tn{grv}(r,\chi)+d^2\dot{E}_\tn{scl}(r,\chi)\ .
\end{equation}

Waveform generation as well as data analysis have been performed 
using Mathematica.\cite{Mathematica} We have sampled the waveform $h(t)$ in the time domain fixing the step $\Delta t=T_\tn{obs}/(2^N-1)$ such that $1/\Delta t=2f_\tn{\rm max}$, where the observation time is $T_\tn{obs}=1$ year and $f_\tn{\rm max}$ corresponds to the ISCO frequency of a test body in the Kerr background. For the EMRI configurations considered, with $M=10^6M_\odot$, $\mu=10M_\odot$ and $\chi=0.9$, this yields to $f_\tn{max}\simeq0.01457$Hz with $\Delta t\simeq 30$ seconds, and a Nyquist frequency of $f_\tn{Ny}=1/(2\Delta t)\simeq0.01663$Hz.
Before applying the Fourier transform to $h(t=n\Delta t)$ for $n=0,\ldots N-1$, we have tapered the waveform with a Tukey window $w[n]$ to avoid boundary effects:
\begin{equation}
	w[n]=\begin{cases}
		\sin\left[\frac{n\pi}{\tau(1-N)}\right]^2 & 0\le n\leq \frac{\tau (N-1)}{2}\ ,\nonumber\\
		1 & \frac{\tau (N-1)}{2}\le n\leq  (N-1)(1-\frac{\tau}{2})\ ,\nonumber\\
		\sin\left[\frac{(1+n-N)\pi}{\tau(1-N)}\right]^2   
		& (N-1)(1-\frac{\tau}{2}) \le n\leq N-1 \ .
	\end{cases}
\end{equation}
The parameter $\tau$ which controls the magnitude of the sinusoidal lobes has been fixed to $\tau=0.05$. We have performed the Fisher matrix analysis without the Tukey window as well and this has led to a mild improvement of a factor $\sim2$ for the errors of the 
waveform parameters.\\

We have computed numerical derivatives for the Fisher matrix using a centered 11-point stencil\,\cite{Milne-ThomsonLouisMelville1951Tcof} for the waveform parameters $\vec{\theta}=(\ln M,\ln \mu,\chi,d,r_0)$ and a centered 9-point stencil for $\vec{\theta}=(\theta_s,\phi_s,\theta_l,\phi_l)$ 
\begin{align}
	\partial_{\theta_i}h(f,\vec{\theta})&=\frac{1}{1260 \delta\theta_i}\vec{P}^{(11)}\cdot \vec{D}^{(11)} + \mathcal{O}(\delta\theta_i^{10})\ ,\nonumber\\
	\partial_{\theta_i}h(f,\vec{\theta})&=\frac{1}{840 \delta\theta_i}\vec{P}^{(9)}\cdot \vec{D}^{(9)} + \mathcal{O}(\delta\theta_i^8)\ ,\label{math:num_derivatives}
\end{align}
where $\vec{P}^{(9)}=(3,-32,168,-672,0,672,-168,32,-3)$, 
$\vec{P}^{(11)}=(-1,25/2,-75,300,-1050,1050,-300,75,-25/2,1)$, 
$\vec{D}^{(8)}=(h_{i-4},h_{i-3},\ldots,h_{i+3},h_{i+4})$, 
$\vec{D}^{(11)}=(h_{i-5},h_{i-4},\ldots,h_{i+4},h_{i+5})$, 
and $h_{i\pm n}=h(fs,\theta_i\pm n\delta\theta_i)$. For $\theta_i\neq0$ ($\theta_i=0$) we set $\delta\theta_i=\theta_i \epsilon_i$ ($\delta\theta_i=\epsilon_i$), 
with the numerical coefficients $\epsilon_i\ll 1$. The difference in the 
derivative order is due to the extreme stability of angular parameters 
which can be treated with a lower order method.

Derivatives with respect to the initial phase and the luminosity distance have an analytic expression since $\partial_{\ln D}h(f)=-h(f)$ and since $\partial_{\Phi_0}$ can be directly applied to the time domain waveform: 
\begin{equation}
	\begin{split}
		\partial_{\Phi_0}h(t,\vec{\theta})=
		-\sqrt{3}{\cal A} F_+ &\sin[2\Phi(t)+2\Phi_0](1+\cos^2\iota)\\
		&-2\sqrt{3}{\cal A}F_\times \cos[2\Phi(t)+2\Phi_0]\cos\iota\  .
	\end{split}
\end{equation}
We have computed Eq.~\eqref{math:num_derivatives} varying the spacings $\epsilon_i$ in order to explore the behavior of the derivatives. 
The choice of $\epsilon_i$ is also relevant for the inversion of the Fisher matrix. Integration over the LISA noise spectral density has been performed using a composite Boole method.\cite{Press:1992zz} 
We have checked that choosing a lower-order method for the derivatives, or a different method for the frequency integral, do not change our 
final results.

\begin{figure}
	\centering
	\includegraphics[width=9cm]{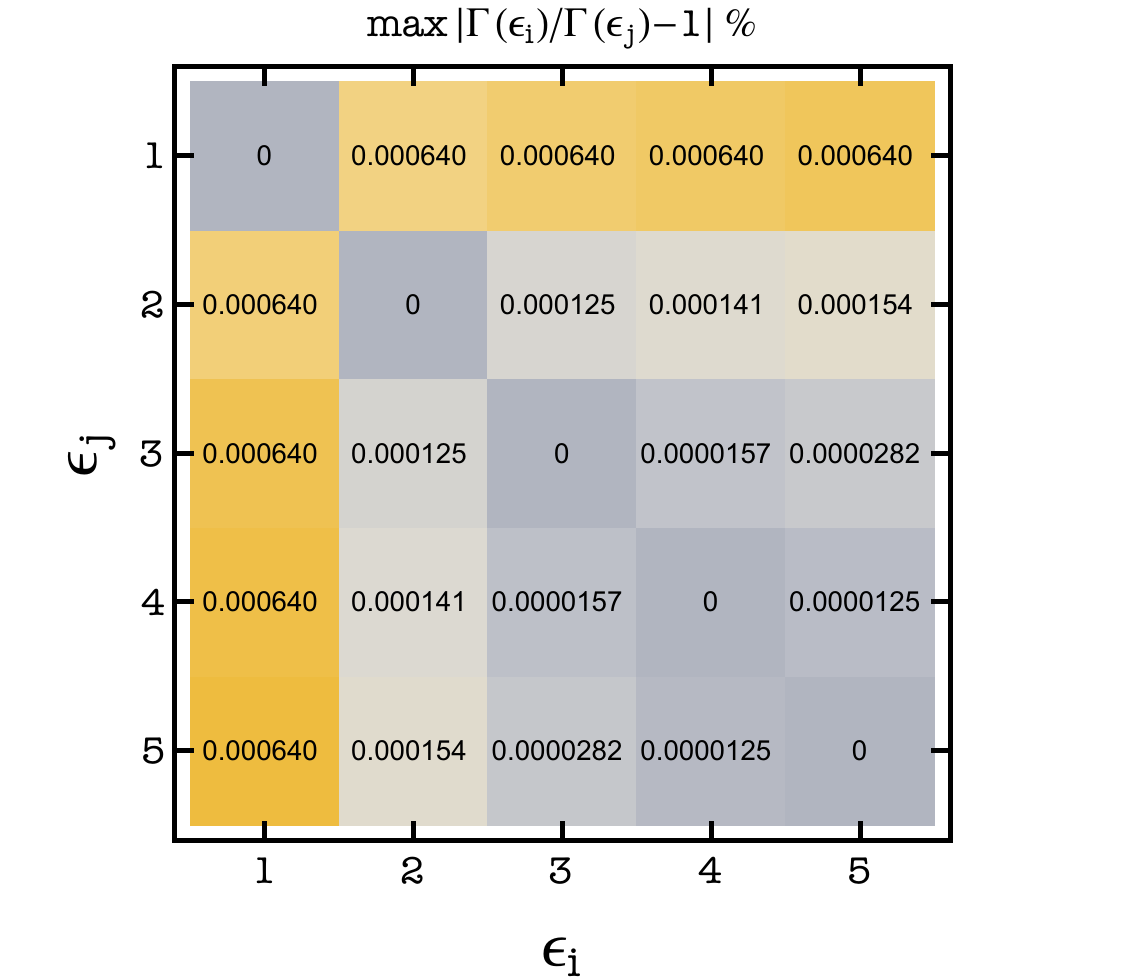}
	\caption{
		\textbf{Maximum relative (percentage) errors between Fisher matrices  computed assuming different values  of the spacing used for the numerical  derivatives.} The Fisher matrices have been computed  for EMRIs with $d=0.2$.}
	\label{fig:fisher accuracy}
\end{figure}

%-------------------------------------------------------------
\textbf{Numerical stability.}
We have performed various checks in order to assess the stability and convergence of our numerical calculations. The correlation matrix in Figure~\ref{fig:fisher accuracy} shows the maximum relative error, $|\Gamma_{kn}(\epsilon_i)/\Gamma_{kn}(\epsilon_j)-1|\times 100$, between Fisher matrices computed for different values of the spacing, where $\epsilon_{i=1,\ldots5}=(10^{-5}, 5\times 10^{-6},10^{-6},5\times 10^{-7},10^{-7})$ for the scalar charge and $\epsilon_{i=1,\ldots5}=(10^{-7}, 5\times 10^{-8}, 10^{-8}, 5\times 10^{-9},10^{-9})$ for the other parameters. Values refer to EMRIs with $d=0.2$, although the other configurations discussed in the paper yield similar results. The Fisher matrices  computed with various $\epsilon_i$ differ less than one part over $10^5$.\\

\begin{figure}
	\centering
	\includegraphics[width=9cm]{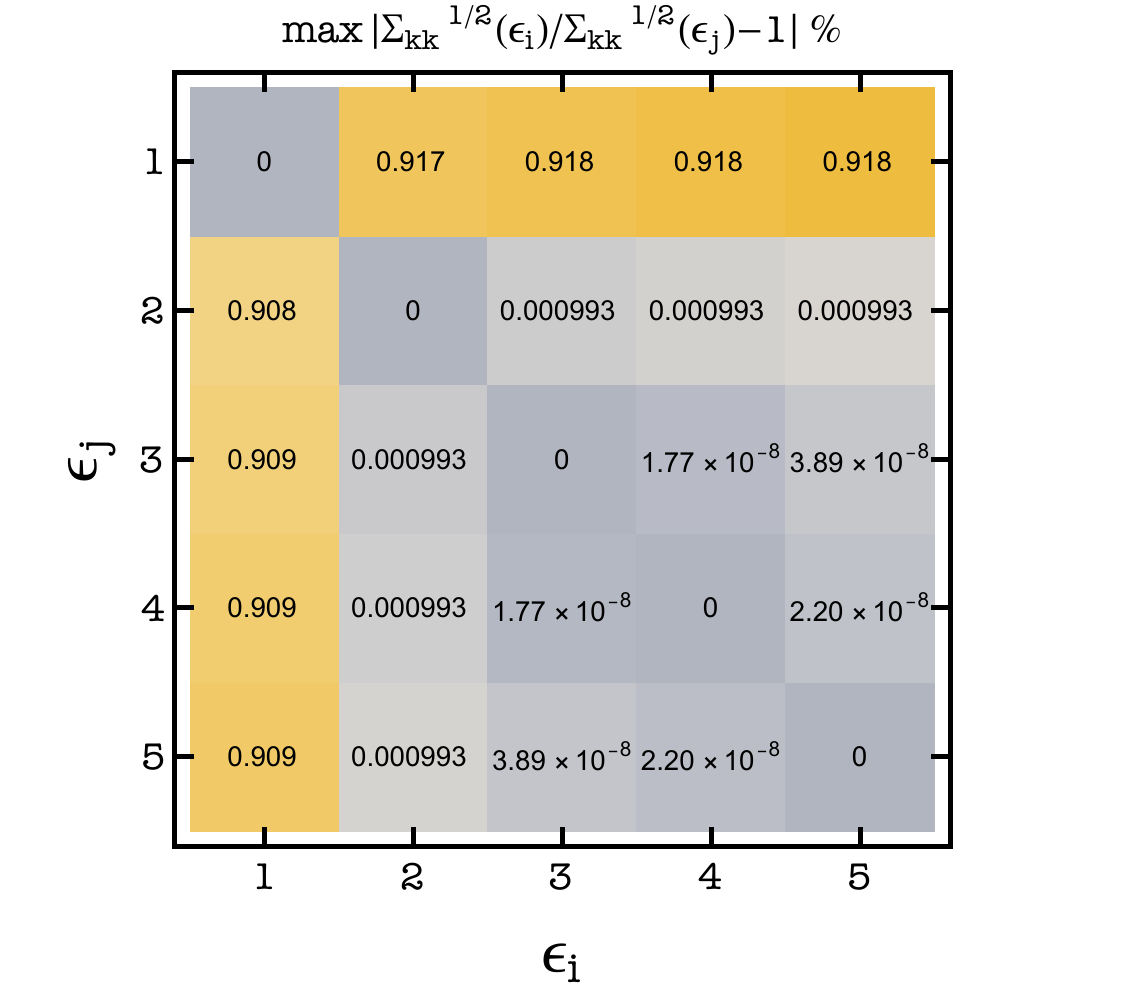}
	\caption{
		\textbf{Maximum relative (percentage) errors between (square root of) the diagonal components of the covariance  matrices, i.e., the parameter errors, obtained with  different values of the shifts considered for  the numerical derivatives.} 
		We consider the same EMRI  configurations shown in Figure~\ref{fig:fisher accuracy}.}
	\label{fig:error accuracy}
\end{figure}

\begin{figure}
	\includegraphics[width=\columnwidth]{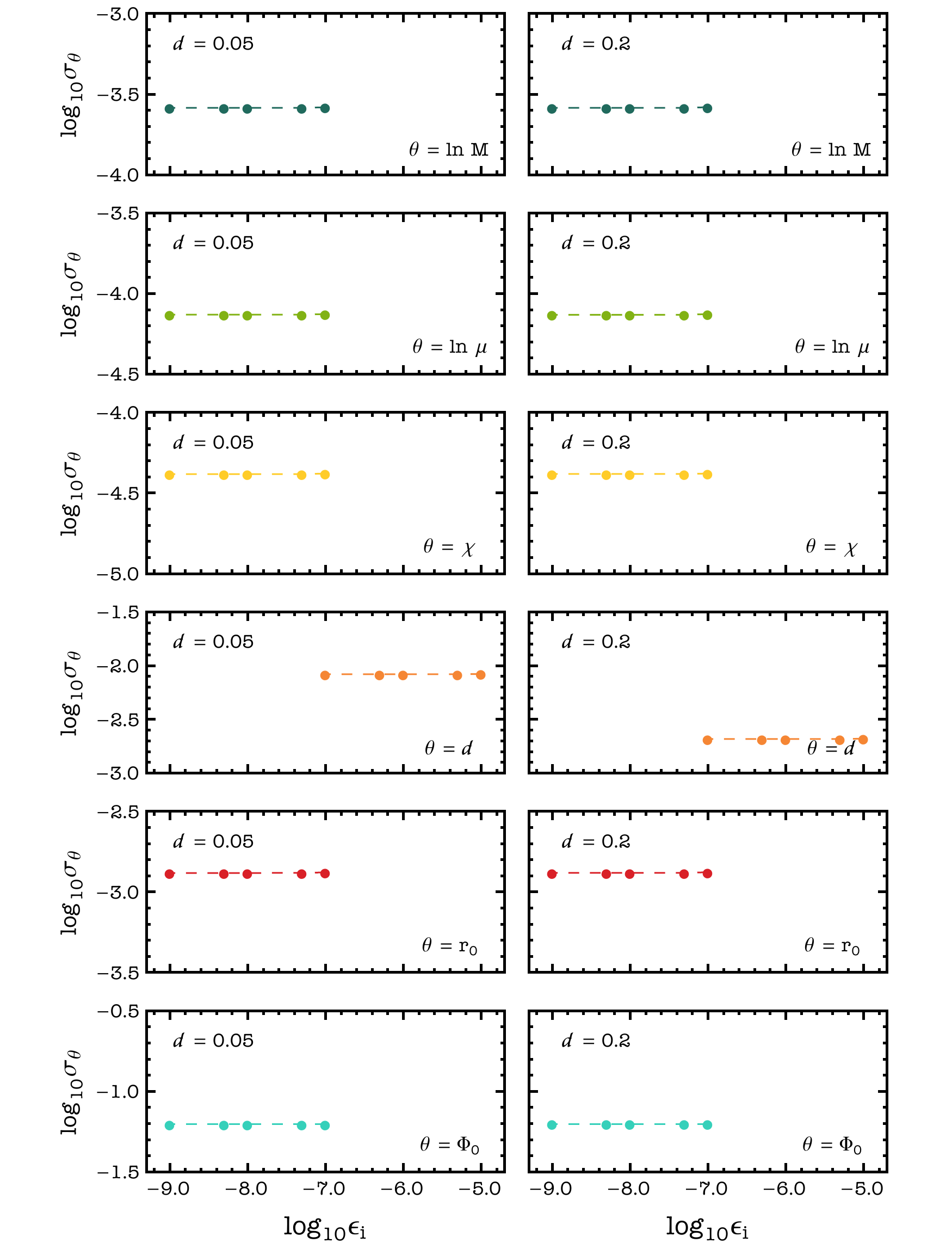}
	\caption{\textbf{Errors on the EMRI parameters as a function of the numerical shift used for  partial derivatives of the Fisher matrix.} We consider binaries with $d=0.05$ (left panel) and $d=0.2$ (right panel)  with SNR of $150$ as observed by LISA one year before the plunge.}
	\label{fig:errors_vs_epsilon}
\end{figure}

In the case of EMRIs, it is well known that computations of the Fisher matrix may lead to ill-conditioned problems, and be plagued by numerical instabilities.\cite{Porter:2009wi,Porter:2015eha} The intrinsic high sensitivity of EMRIs to small perturbations of the system reflects their ability to provide exquisite constraints on the waveform parameters and hence it is a blessing in disguise. This large amount of information is reflected on the magnitude of the elements of the Fisher matrix, although not all of them have the same size. For all the EMRI configurations analysed, the components of ${\bf \Gamma}$ corresponding to $(M,\mu, \chi, r_0,d)$ are always predominant with respect to the rest of the waveform parameters, with differences of several orders of magnitude.
These differences lead to large condition numbers, i.e., the ratios between the largest to the smallest of the Fisher eigenvalues $\lambda_{i=1,\ldots 11}$, which in our case are of the order $\kappa={\rm max}(\lambda_i)/{\rm min}(\lambda_i)\sim\mathcal{O}(10^{14})$. 
Nonetheless, the high precision computation we perform leads to stable inversion of the Fisher matrices, independent of the numerical shift. 
The correlation matrix in Figure~\ref{fig:error accuracy} shows the maximum relative error between the square root of the diagonal components of the covariance matrices, i.e., the parameter errors, computed by inverting the Fisher matrix derived with different $\epsilon_i$, for the $d=0.2$ case. We see from this analysis that a large fraction of errors agree with each other to well below the $1\%$ level, and the few remaining cases feature differences of at most $1\%$. Correlation coefficients are also very stable, with differences smaller than $1\%$.
For the sake of completeness we show in Figure~\ref{fig:errors_vs_epsilon} the errors on masses, spins, initial phase and radius, and on the charge, for two EMRI configurations with $d=0.05$ and $d=0.2$. The various panels demonstrate again the stability of our error calculations. The constraints inferred for $(M,\mu,\chi,r_0,\Phi_0)$ are also consistent with previous results on LISA parameter estimation, performed with different approaches and GW templates.\cite{Huerta:2011kt,Huerta:2011zi,Piovano:2021iwv}

\begin{figure}
	\includegraphics[width=\columnwidth]{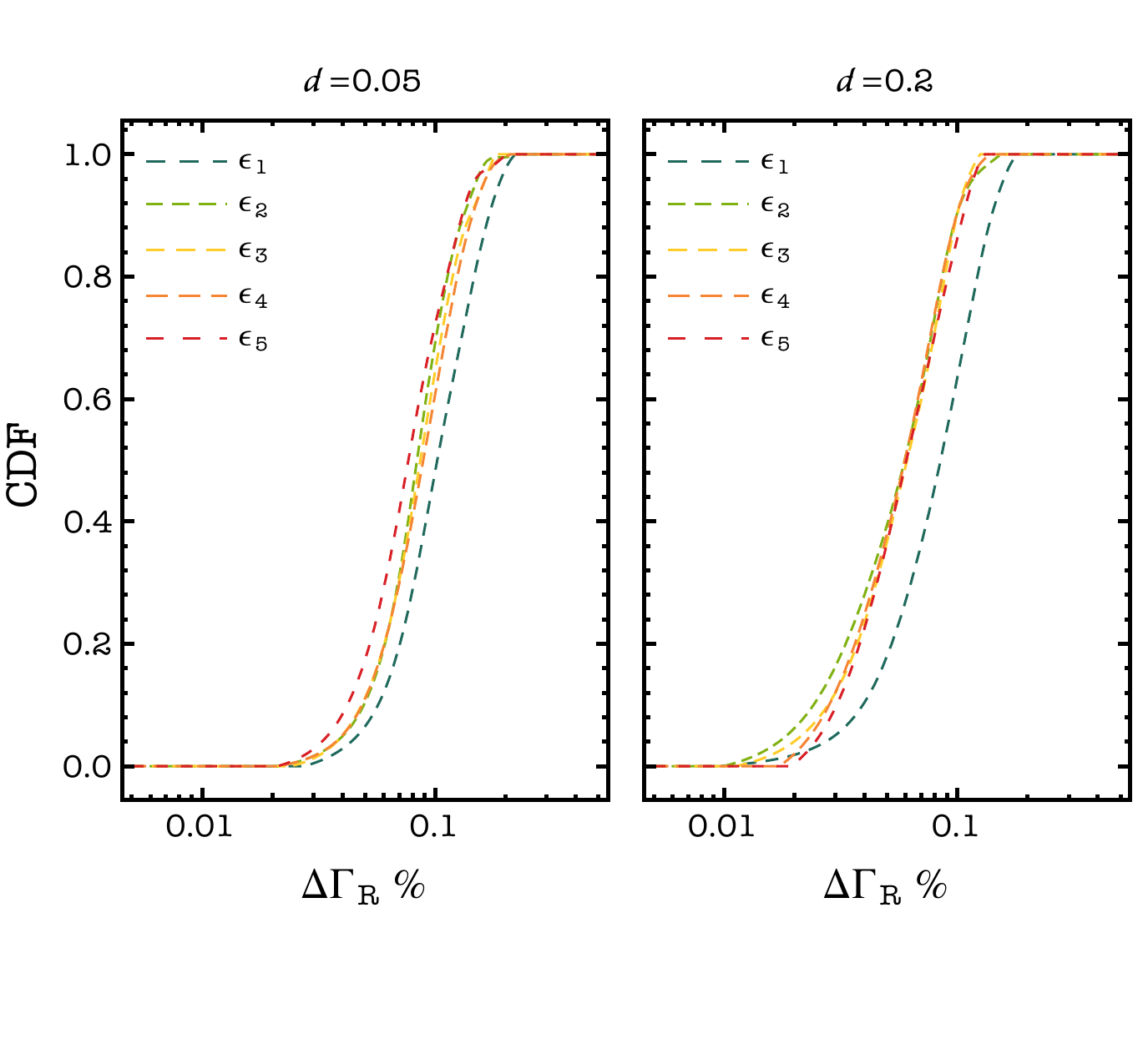}
	\caption{
		\textbf{Cumulative distribution for the maximum relative error between unperturbed and perturbed Fisher matrices with elements  shifted by random numbers drawn from a uniform distribution.} 
		Colored curves refer to ${\bf \Gamma}$ computed with a different choice of the numerical derivative spacing. The uniform distribution is defined within
		$[-10^{-3},10^{-3}]$.}
	\label{fig:stability}
\end{figure}

We have further checked the stability of our calculations perturbing the Fisher matrices and studied the results of the inversion. Following previous work,\cite{Speri:2021psr} we build a matrix ${\bf R}$ with the same dimensionality as ${\bf \Gamma}$, with entries randomly drawn from a uniform distribution $U\in[-10^{-3},10^{-3}]$. We then compute the inverse $({\bf \Gamma}+{\bf R})^{-1}$ and determine the maximum relative error with respect to the unperturbed configuration, $\Delta {\bf \Gamma_{R}}={\rm max}({\bf \Gamma}+{\bf R})^{-1}/{\bf \Gamma}^{-1}-{\bf I})$. This procedure has been iterated 100 times, in order to build up statistics for the maximum error. Figure~\ref{fig:stability} shows the cumulative distribution of $\Delta {\bf \Gamma_{R}}$ for two EMRIs with different values of the charge, as a function of the derivative spacing. The picture suggests that the calculations are very stable, as in the majority of cases more than $90\%$ of the population has $\Delta {\bf \Gamma_{R}}\lesssim0.1\%$. We find similar results for all the other binaries analysed.

\begin{figure}
	\centering
	\includegraphics[width=0.65\columnwidth]{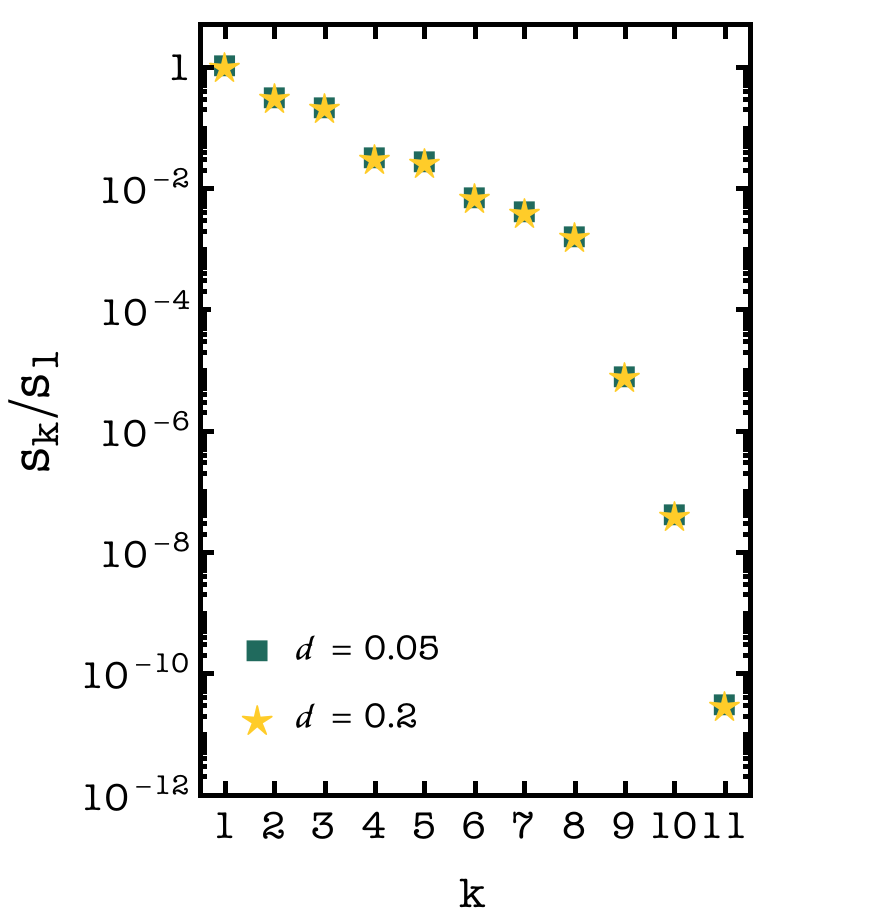}
	\caption{
		\textbf{Singular values normalised to their largest component for Fisher matrices with different values of the scalar charge and secondary mass.} 
		Calculations have been performed assuming $\epsilon=10^{-5}$ for $\partial_d$ and $\epsilon=10^{-7}$ for the rest of the derivatives, although varying the spacing does not lead to significant changes in this figure.}
	\label{fig:singular_values}
\end{figure}

%-------------------------------------------------------------
\textbf{Singular value decomposition.}
Beside direct inversion of the Fisher matrix, we have derived the covariance on $\vec{\theta}$ also applying a truncated singular value (SVD) decomposition approach\cite{Pai:2012mv,Wade:2013hoa,Vallisneri:2007ev} on ${\bf \Gamma}$.
Any matrix ${\bf A}\in \mathbb{R}^{n\times n}$ can be written in terms of its SVD as
\begin{equation}
	{\bf A}= {\bf U} {\bf S} {\bf V}^\tn{T}\ ,
\end{equation}
where ${\bf U}$ and ${\bf V}$, with columns called {\it singular vectors}, are orthonormal matrices 
${\bf U} {\bf U}^\tn{T}={\bf U}^\tn{T}{\bf U}={\bf V}^\tn{T}{\bf V}={\bf V} {\bf V}^\tn{T}={\bf I}$, and 
\begin{equation}
	{\bf S}={\rm diag}(s_1,s_2,\ldots s_n)\ ,
\end{equation}
where $\sigma_i\ge 0$ are the {\it singular values}. From the SVD computing the inverse of ${\bf A}$ is also straightforward, since 
\begin{equation}
	{\bf A}^{-1}={\bf V}{\bf S}^{-1}{\bf U}^\tn{T}\,.
\end{equation}
The singular vectors define the dimensions of the variance of the data, ordered in magnitude such that the first one is the largest, while the rank $r$ of the initial matrix correspond to the number of non-zero $\sigma_i$. Given the form of ${\bf S}$ we can also recast ${\bf A}$ as 
\begin{equation}
	{\bf A}= {\bf U} {\bf S} {\bf V}^\tn{T}=\sum_{i=1}^r\sigma_i{\bf u}_i{\bf v}^\tn{T}_i\ ,
\end{equation}
which tells that the original matrix can be decomposed into a sum of rank-$1$ layers with the first contributing the most. Note that before applying the SVD, we normalize the Fisher matrices to the variance of their components, namely ${\bf \Gamma}/{\bf N}$, where ${\bf N}={\rm diag}({\bf \Gamma}) \otimes {\rm diag}({\bf \Gamma})$, to remove differences between the parameters given by their physical scales.
Although, as discussed above, the accuracy of our calculations guarantee an extremely stable inversion, we have also applied the SVD to obtain the covariance on the source parameters. In particular we have removed the singular pieces of the Fisher by zeroing values of ${\bf S}^{-1}$ which are very large, i.e., corresponding to small $s_i$. 
Figure \ref{fig:singular_values} shows a scree plot of the $d_i$ normalised to the first one, for a binary with $d=0.05$ and $d=0.2$ (other configurations feature the same behavior, regardless of the choice of $\epsilon_i$ as well). 

\begin{figure}
	\centering
	\includegraphics[width=\columnwidth]{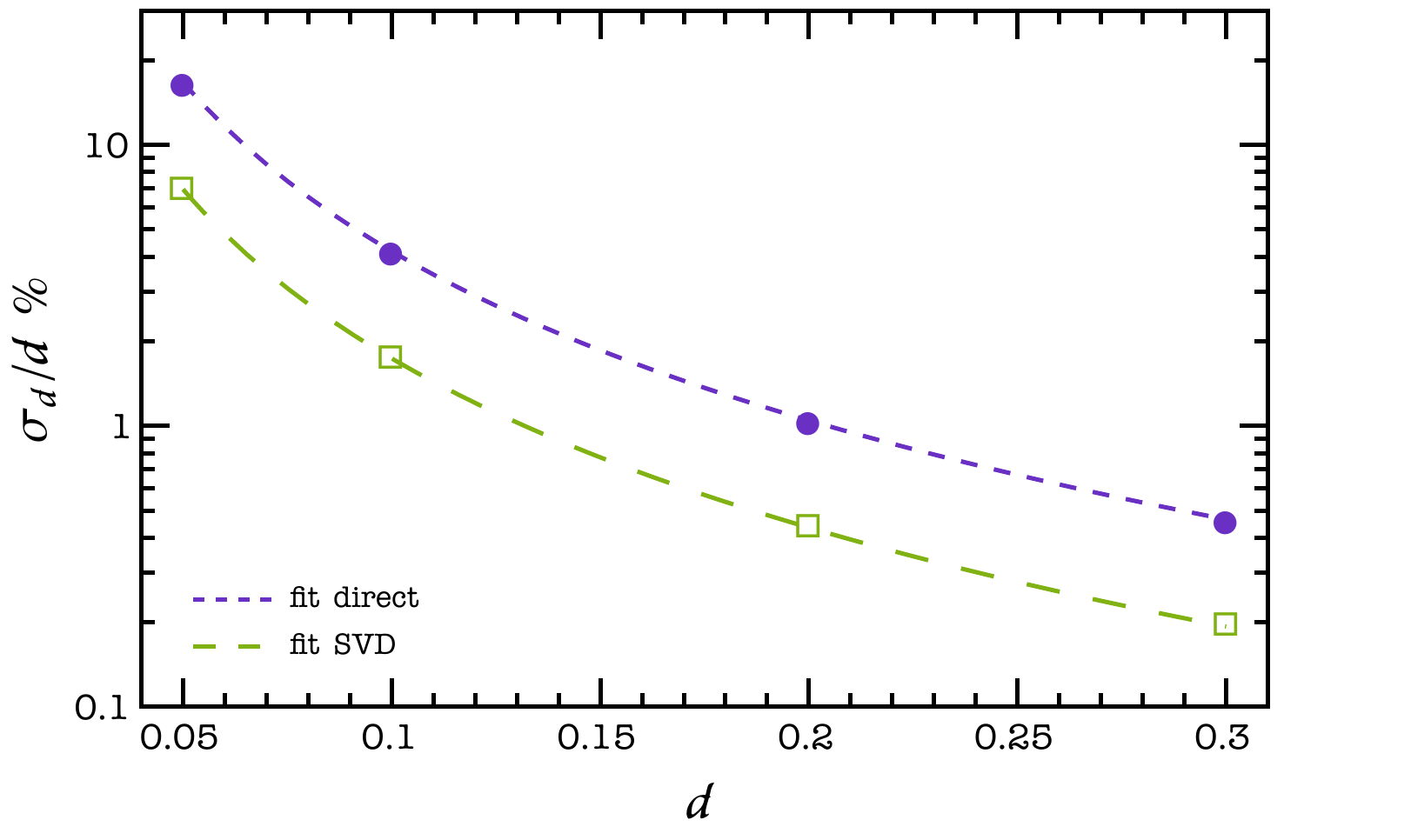}
	\caption{
		\textbf{Relative error on the scalar charge in the cases of direct inversion and singular value decomposition.}
		Filled and empty dots refer to errors computed to direct inversion  or truncated SVD approach. Dashed lines identify analytic fit for  the errors $\sigma_d=\beta/d$, with $\beta\simeq4.18\times 10^{-4}$, and $\beta\simeq1.74\times 10^{-4}$ for the two methods considered  to invert the  Fisher matrices.}
	\label{fig:singular_errors}
\end{figure}

The trend of the singular values exhibits a clear drop around the last component where a steep decrease in the magnitude of the $s_i$ appears. Here we compute ${\bf \Gamma}^{-1}$ zeroing the last term of ${\bf S}^{-1}$. This procedure improves in general the errors, as it removes unmeasurable linear combinations of parameters from the Fisher matrix.\cite{Vallisneri:2007ev} Indeed, as shown in Figure~\ref{fig:singular_errors} in which we compare relative errors on the scalar charge using direct inversion and the SVD approach, the latter yields a significant reduction of the errors. In the results shown in the main part of the paper we have reported the more conservative, direct inversion results, but the SVD analysis suggests that LISA might be able to measure scalar charge with even higher precision.

%\bibliographystyle{naturemag}
%\bibliography{references}

\end{document}